\documentclass[fleqn,twoside]{article}
\usepackage{amsmath,amssymb,espcrc2}

\usepackage{graphicx}
\usepackage[figuresright]{rotating}

\newcommand{\url}[1]{{\tt #1}}

\newcommand{\gmt}{$(g-2)_\mu$}
\newcommand{\bsg}{BR($b \to s \gamma$)}
\newcommand{\btn}{BR($B_u \to \tau \nu_\tau$)}
\newcommand{\MW}{$m_W$}
\newcommand{\mh}{m_h}

\graphicspath{{figs/}}

\title{Predictions for Supersymmetric Particle Masses 
using Indirect Experimental and Cosmological Constraints} 

\author{
O.~Buchmueller\address[CERN]
   {CERN, CH-1211 Geneve 23, Switzerland},
R.~Cavanaugh\address[FNAL]
   {Fermi National Accelerator Laboratory, P.O. Box 500, 
    Batavia, Illinois 60510, USA}\hbox{$^{\rm ,}$}\address[UIC]
   {Physics Department, University of Illinois at Chicago, Chicago, 
    Illinois 60607-7059, USA},
A.~De Roeck\addressmark[CERN]\hbox{$^{\rm ,}$}\address[Antwerpen]
   {Antwerp University, B-2610 Wilrijk, Belgium},
J.R.~Ellis\addressmark[CERN],
H.~Fl\"acher\addressmark[CERN],
S.~Heinemeyer\address[Santander]
   {Instituto de F\'{\i}sica de Cantabria (CSIC-UC), 
    E--39005 Santander, Spain},
G.~Isidori\address[Isidori]
   {INFN, Laboratori Nazionali di Frascati, Via E. Fermi 40, 
    I--00044 Frascati, Italy}\hbox{$^{\rm ,}$}\address[Pisa]
   {Scuola Normale Superiore, Piazza dei Cavalieri 7, I-56126 Pisa, Italy},
K.A.~Olive\address[Minnesota] 
   {School of Physics and Astronomy, University of Minnesota, Minneapolis,
    Minnesota 55455, USA}, 
P.~Paradisi\address[Paradisi]
   {Physik-Department, Technische Universit\"at M\"unchen, 
  D-85748 Garching, Germany},
F.J.~Ronga\address[ETHZ]
   {Institute for Particle Physics, ETH Z\"urich, CH-8093 Z\"urich, 
   Switzerland},
G.~Weiglein\address[Durham]
   {IPPP, University of Durham, Durham DH1 3LE, U.K.}
}
       
\begin{document}

\begin{abstract}
In view of the imminent start of the LHC experimental programme, we use the
available indirect experimental and cosmological information to estimate the
likely range of parameters of the constrained minimal supersymmetric extension
of the Standard Model (CMSSM), using a Markov-chain Monte Carlo (MCMC)
technique to sample the parameter space. The 95\% confidence-level area in
the $(m_0, m_{1/2})$ plane of the CMSSM lies largely within the region that
could be explored with 1~fb$^{-1}$ of integrated luminosity at 14~TeV,
and much of the 68\% confidence-level area  lies within the region that
could be explored with 50~pb$^{-1}$ of integrated luminosity at
10~TeV. A same-sign dilepton signal could well be visible in most of the
68\% confidence-level area with 1~fb$^{-1}$ of integrated luminosity at
14~TeV. We discuss the sensitivities of the
preferred ranges to variations in the most relevant indirect
experimental and cosmological constraints and also to deviations from
the universality of the supersymmetry-breaking contributions to the
masses of the Higgs bosons.

\bigskip
\begin{center}
CERN-PH-TH/2008-181, DCPT/08/118, FTPI-MINN-08/33, IPPP/08/59, UMN-TH-2715/08
\end{center}
\vspace{-0.5cm}
\end{abstract}

\maketitle

\section{Introduction}
\label{sec:intro}

Supersymmetry (SUSY)~\cite{Nilles:1983ge,Haber:1984rc,Barbieri:1982eh} 
is one of the most highly favoured extensions of the
Standard Model (SM), and is often considered to be a prime candidate for
discovery at the Large Hadron Collider (LHC). With the start of
experiments at the LHC now 
becoming imminent, it is natural and topical to make the best possible
assessment of the likelihood that the LHC will indeed 
discover SUSY, based on the best available experimental,
phenomenological and cosmological information. Most of the current
constraints on possible physics beyond the SM are negative, in the sense
that they reflect the agreement of data with the SM, and set only lower
limits on the possible masses of supersymmetric particles~\cite{pdg}.
Examples are direct constraints such as lower limits on specific sparticles,
e.g., the chargino, and indirect constraints such as the lower limit on the
possible mass of a SM-like Higgs boson~\cite{Barate:2003sz,Schael:2006cr}.

However, there are two observational constraints that, within the
context of SUSY, 
may be used also to set upper limits on the possible masses of supersymmetric
particles, since they correspond to measurements that cannot be explained by
the SM alone. These hints for new physics are the anomalous magnetic moment of
the muon, \gmt, which appears to differ by over three standard deviations from
the best SM calculation based on low-energy $e^+ e^-$
data~\cite{Bennett:2006fi,Moroi:1995yh,Czarnecki:2001pv,Davier:2007ua,Miller:2007kk,Jegerlehner:2007xe,Passera:2008jk},
and the density of cold dark matter, $\Omega_{\rm CDM}$~\cite{Dunkley:2008ie},
which has no possible origin within the SM.
Each of these discrepancies has many possible interpretations, of which
SUSY is just one. Nevertheless, given the strong motivations for
SUSY, which include the naturalness of the mass hierarchy and
grand unification, as well as the existence of a plausible candidate for the
astrophysical cold dark matter, it is natural to ask what \gmt\ and
$\Omega_{\rm CDM}$ may imply for the parameters of supersymmetric models.
Any such analysis should also take into account the constraints imposed by
precision measurements of electroweak observables (EWPO) and $B$-physics
observables (BPO) such as \bsg, where most observables 
agree quite well with the SM.

In this paper we revisit the indirect information on supersymmetric model 
parameters obtainable in
the light of these experimental, phenomenological and cosmological constraints,
using a Markov-chain Monte Carlo (MCMC) approach, see, e.g.,
Ref.~\cite{Allanach:2007qj} and references therein. This is practical only
in simplified versions of the minimal supersymmetric extension of the Standard
Model (MSSM), in which some universality relations are imposed on the
soft SUSY-breaking parameters. Initially, we work in the framework of
the constrained MSSM (CMSSM), in which the scalar and gaugino mass
parameters, $m_0$, $m_{1/2}$, and the trilinear coupling $A_0$ are each
assumed to be equal at the input GUT scale. Furthermore as low-energy
parameter we have $\tan\beta$, the ratio of the two vacuum expectation
values of the two Higgs doublets. 
At the end we also comment on the possible changes in our results if the
common soft SUSY-breaking contribution to the Higgs scalar
masses-squared, $m_H^2$, is allowed 
to differ from those of the squarks and sleptons, the single-parameter
non-universal Higgs model or NUHM1.

There have been many previous studies of the CMSSM parameter
space~\cite{Djouadi:2001yk,deBoer:2001nu,deBoer:2003xm,eos2,baer,arndut,eoss,baer2,lahnan,nath,munoz,Belanger:2004ag,Ellis:2003si,Ellis:2004tc,Ellis:2005tu,Ellis:2006ix,Ellis:2007aa,Ellis:2007ss,Baltz:2004aw,Allanach:2005kz,Allanach:2006jc,Allanach:2006cc,Allanach:2007qk,Feroz:2008wr,deAustri:2006pe,Roszkowski:2006mi,Roszkowski:2007fd,Roszkowski:2007va,Heinemeyer:2008fb,Bechtle:2004pc,Lafaye:2007vs,Buchmueller:2007zk,Allanach:2008iq}, 
including estimates of the sparticle masses, and a number of these have
used MCMC
techniques~\cite{Baltz:2004aw,Allanach:2005kz,Allanach:2006jc,Allanach:2006cc,Allanach:2007qk,Feroz:2008wr,deAustri:2006pe,Roszkowski:2006mi,Roszkowski:2007fd,Roszkowski:2007va}.
These have been used to 
extract the preferred values for the CMSSM parameters using low-energy
precision data, bounds from astrophysical observables and flavour-related
observables.
These analyses differ in the precision observables that have been considered,
the level of sophistication of the theory predictions that have been used, and
the way the statistical analysis has been performed. 

Here we use the MCMC technique to sample efficiently the SUSY parameter space,
and thereby construct the $\chi^2$ probability function,
$P(\chi^2,N_{\rm dof})$. This accounts correctly for the number of degrees of
freedom, $N_{\rm dof}$, and thus represents a quantitative measure for the
quality-of-fit.  Hence $P(\chi^2,N_{\rm dof})$ can be used to estimate the
absolute probability with which the CMSSM describes the experimental data. Our
probabilistic treatment is explained in detail in Sec.~\ref{sec:mpfit}.


Many previous analyses found evidence for a relatively low SUSY mass
scale in  the stau-coannihilation region,
e.g.,~\cite{Ellis:2007aa,Buchmueller:2007zk,trotta}, and a 
mild preference for  $\tan\beta \sim 10$ was found
in~\cite{Buchmueller:2007zk}. A comparison of Bayesian analyses 
yielding varying results under
different assumptions was made in~\cite{Allanach:2008iq}. 
Some differences between analyses may also be
traceable to the treatments of the \bsg\ and \gmt\ measurements,
which we discuss in some detail below.


Our main objectives in this paper are threefold.
One is to discuss explicitly the prospects for discovering sparticles 
in early LHC running, another is to discuss the
robustness of the fit results by analyzing the
implications of relaxing the constraints due to \gmt, 
\bsg, $\Omega_{\rm CDM}$ and other observables, 
and the third is to discuss the extension of the CMSSM results to the NUHM1,
in which an extra parameter is introduced that allows a common degree
of non-universality for the two Higgs multiplets.

We find that the 95\%~C.L.\ area in the $(m_{1/2}, m_0)$ plane of the
CMSSM lies largely within the region that could be explored with
1~fb$^{-1}$ of integrated LHC luminosity at 14~TeV
in a single experiment, and that much of the 68\%~C.L.\
area  lies within the region that could be explored with
50~pb$^{-1}$ of integrated luminosity at 10~TeV (the projected initial
LHC collision energy). A same-sign dilepton
signal could well be visible in the 68\%~C.L.\ area with 1~fb$^{-1}$ of
integrated luminosity at 14~TeV, and the lightest Higgs boson might
also be detectable in squark decays with 2~fb$^{-1}$ of integrated
luminosity at 14~TeV. We find that removing the $\Omega_{\rm CDM}$
constraint has little effect on the preferred regions of the CMSSM
parameter space in the $(m_0, m_{1/2})$, $(\tan \beta, m_{1/2})$, and
$(\tan \beta, m_0)$ planes, apart from expanding the range of $m_0$,
particularly for $\tan \beta \sim 10$. On the other hand, rescaling the
present error in \gmt\ may have quite an important effect: the
preferred ranges in $m_{1/2}$ and $m_0$ would expand quite significantly
if the error on the present experimental discrepancy with the SM were to
be increased. Conversely, if this error could be reduced, e.g., by a
more precise measurement of \gmt\ and/or a more refined theoretical
estimate within the SM, the predictions for sparticle masses could be
significantly improved. We also discuss the effects of possible
variations in the errors in \bsg\ and other observables. Finally, we show
that our results would not be greatly changed 
in the NUHM1: we leave a more complete study of the NUHM1 and the
NUHM2 (in which the masses of the two Higgs multiplets are
independently non-universal) for future work. 

\section{Multi-parameter Fit to Experimental Observables}
\label{sec:mpfit}

\begin{table*}[htb!]
\renewcommand{\arraystretch}{1.3}
\begin{center}
\begin{tabular}{|c|c|c|c|c|} \hline
Observable & Th.\ Source & Ex.\ Source & Constraint & Add.\ Th.\ Unc. \\ 
\hline \hline
\MW [GeV] 
    &\cite{Heinemeyer:2006px,Heinemeyer:2007bw}   
    &\cite{verzocchi} & $80.399 \pm 0.025$  & 0.010 \\ 

\hline
$ a_{\mu}^{\rm exp} - a_{\mu}^{\rm SM}$ 
    &\cite{Moroi:1995yh,Degrassi:1998es,Heinemeyer:2003dq,Heinemeyer:2004yq} 
    &\cite{Bennett:2006fi,Davier:2007ua,Hertzog:2007hz} 
                                        &$(30.2 \pm 8.8)\times10^{-10}$ 
                                        &$2.0\times10^{-10}$ \\ 
\hline
$\mh$ [GeV] 
    & \cite{Degrassi:2002fi,Heinemeyer:1998np,Heinemeyer:1998yj,Frank:2006yh} 
    & \cite{Barate:2003sz,Schael:2006cr} & $> 114.4$ (see text) & 3.0  \\ 

\hline
BR$_{\rm b \to s \gamma}^{\rm exp}/ {\rm BR}_{\rm b \to s \gamma}^{\rm SM} $
    &\cite{Misiak:2006zs,Ciuchini:1998xy,Degrassi:2000qf,Carena:2000uj,D'Ambrosio:2002ex}
    &\cite{hfag}       
    &$1.117 \pm 0.076_{\rm exp} \pm 0.082_{\rm th(SM)}$  & 0.050 \\ 

\hline
\hline

$m_t$ [GeV] 
    &\cite{Heinemeyer:2006px,Heinemeyer:2007bw}   
    &\cite{Group:2008vn} & $172.4 \pm 1.2$    & -- \\ 

\hline
$\Omega_{\rm CDM} h^2$ 
    &\cite{Belanger:2006is,Belanger:2001fz,Belanger:2004yn} &\cite{Dunkley:2008ie}  
        &$0.1099 \pm 0.0062$ & 0.012 \\ 

\hline
BR$(B_{s} \to \mu^{+} \mu^{-})$ 
    &\cite{Isidori:2001fv,BCRS,Isidori:2006pk,Isidori:2007jw} &\cite{hfag}
    &$ < 4.7 \times 10^{-8}$   & $0.02\times10^{-8}$ \\ 

\hline 
\hline

BR$_{\rm B \to \tau\nu}^{\rm exp}/ {\rm BR}_{\rm B \to \tau\nu}^{\rm SM} $
    &\cite{Isidori:2006pk,Isidori:2007jw,Akeroyd:2003zr} &\cite{Aubert:2004kz,paoti,Latticefb}    
    &  $1.15 \pm 0.40_{\rm [exp+th]}$  & -- \\ 

\hline
${\rm BR}({B_d \to \mu^+ \mu^-})$
    & \cite{Isidori:2001fv,BCRS,Isidori:2006pk,Isidori:2007jw} &\cite{hfag} 
    & $ <2.3 \times 10^{-8}$ & $0.01 \times 10^{-9}$\\

\hline
${\rm BR}_{B \to X_s \ell \ell}^{\rm exp}/{\rm BR}_{B \to X_s \ell \ell}^{\rm SM}$
    & \cite{Bobeth:2004jz}&\cite{hfag,Huber:2005ig}
    & $0.99 \pm 0.32$ & -- \\

\hline
BR$_{K \to \mu \nu}^{\rm exp}/{\rm BR}_{K \to \mu \nu}^{\rm SM}$
    & \cite{Isidori:2006pk,Akeroyd:2003zr} &\cite{Antonelli:2008jg}
    & $1.008 \pm 0.014_{\rm [exp+th]}$   & -- \\

\hline
BR$_{K \to \pi \nu \bar{\nu}}^{\rm exp}/{\rm BR}_{K \to \pi \nu \bar{\nu}}^{\rm SM}$
    & \cite{Buras:2000qz}&\cite{Artamonov:2008qb} 
    & $ < 4.5 $ & -- \\

\hline
$\Delta M_{B_s}^{\rm exp}/\Delta M_{B_s}^{\rm SM}$
    & \cite{Buras:2000qz} &\cite{Bona:2007vi}
    & $1.11 \pm 0.01_{\rm exp} \pm 0.32_{\rm th(SM)}$ & -- \\

\hline
$\frac{(\Delta M_{B_s}^{\rm exp}/\Delta M_{B_s}^{\rm SM})}{
(\Delta M_{B_d}^{\rm exp}/\Delta M_{B_d}^{\rm SM})}$
    & \cite{Isidori:2001fv,BCRS,Isidori:2006pk,Isidori:2007jw}
&\cite{hfag,Bona:2007vi}
    & $1.09 \pm 0.01_{\rm exp} \pm 0.16_{\rm th(SM)} $  & -- \\

\hline
$\Delta \epsilon_K^{\rm exp}/\Delta \epsilon_K^{\rm SM}$
    & \cite{Buras:2000qz} &\cite{Bona:2007vi}
    & $0.92 \pm 0.14_{\rm [exp+th]}$ & -- \\
\hline

\end{tabular}
\caption{List of experimental constraints used in this work in addition
  to the electroweak observables listed in \cite{Buchmueller:2007zk}.
  The top part of the table shows observables that are very sensitive to
  the MSSM parameter space, the middle part lists observables with
  updated measurements compared   to~\cite{Buchmueller:2007zk} while the
  bottom part lists additional experimental constraints.
  The values and errors shown are the current best understanding of these
  constraints. The rightmost column displays additional theoretical
  uncertainties taken into account when implementing these constraints
  in the MSSM. 
  \label{tab:constraints}}  
\end{center}
\end{table*}

Important observables used in our analysis are listed in
Tab.~\ref{tab:constraints}. Some of the EWPO that are included in the
analysis have not been listed in the Table, 
because they did not change since the analysis carried out 
in~\cite{Buchmueller:2007zk}; their details can be found 
there.

The deviation of \gmt\ from the SM prediction by more than $3\,\sigma$  can
be easily accommodated within the (C)MSSM by choosing appropriately
the sign of the Higgs supermultiplet mixing parameter, $\mu$:
sign($\mu$) = sign($a_{\mu}^{\rm exp} - a_{\mu}^{\rm SM}$). 
Consequently, we analyze
in detail the case $\mu > 0$, and discuss the $\mu < 0$ case only briefly.


The central value of the \bsg\ constraint has changed slightly because of new
experimental results: the data/SM ratio in Tab.~\ref{tab:constraints}
corresponds to the HFAG average \bsg$=(3.52 \pm 0.24)\times
10^{-5}$~\cite{hfag} and to the NNLO SM calculation, 
\bsg$=(3.15 \pm 0.23)\times 10^{-5}$~\cite{Misiak:2006zs}
(both values refer to the inclusive rate with $E_\gamma > 1.6$~GeV).
Despite some interesting recent attempts 
to improve the SM prediction of \bsg\ (see, e.g.,
Refs.~\cite{Becher:2006pu,Gambino:2008fj,Ligeti:2008ac,Misiak:2008ss}  
and references therein), following Ref.~\cite{Misiak:2008ss}
we still consider the above NNLO value 
as the most reliable SM estimate. 
As compared to \cite{Buchmueller:2007zk}, we have reduced the additional
theoretical error in the calculation of the SUSY contribution, for the
following reasons. 
First, it should be noted that all non-perturbative uncertainties cancel
out in the SUSY/SM ratio. Secondly, data force the deviations from the
SM to be small in 
\bsg, so the SUSY/SM ratio can be computed to a relatively high degree of 
accuracy~\footnote{There are two exceptional cases where the theoretical 
uncertainties of the SUSY amplitude can be large: 
i)~the SUSY amplitude is about twice the SM one (SUSY/SM$\sim -2$)
yielding a \bsg\ rate close to the SM value  
ii)~the overall SUSY contribution is small 
because of cancellations among independent large terms.
Case i) is excluded by the $B\to X_s\ell^+\ell^-$
constraints~\cite{Gambino:2004mv,Hurth_new} 
that we take into account in our numerical analysis. We deal with case ii) by
implementing in our code the leading NLO SUSY contributions, that are
known within the MFV framework~\cite{Bobeth:1999ww,Ciuchini:1998xy}.}.
A conservative $15\%$ error on the $b\to s\gamma$ SUSY amplitude
corresponds to less than $5\%$ in the BR$_{\rm b \to s \gamma}^{\rm
  SUSY}/ {\rm BR}_{\rm b\to s\gamma}^{\rm SM}$ ratio in the region where
this does not deviate from unity by more than 30\%.
The \bsg\
constraint, as well as all the other flavour-physics constraints
listed in Tab.~\ref{tab:constraints}, have been implemented using
the code developed in Refs.~\cite{Isidori:2006pk,Isidori:2007jw}.
 This includes the leading NLO QCD
corrections to the supersymmetric contributions~\cite{Ciuchini:1998xy}
and a complete  resummation of all the relevant large $\tan\beta$
effects beyond the lowest 
order~\cite{Degrassi:2000qf,Carena:2000uj,D'Ambrosio:2002ex}.
More recent public codes for the evaluation of 
\bsg\ in the CMSSM have been presented 
in Ref.~\cite{Degrassi:2007kj,Mahmoudi:2008tp}.

A significant $B$-physics constraint arises 
also from BR($B\to\tau\nu$), which represents a powerful probe of the
$(m_{H^{\pm}},\tan\beta)$ plane
\cite{Isidori:2006pk,Isidori:2007jw}. However, at present both
experimental and theoretical uncertainties prevent us from fully
exploiting the potential sensitivity of this observable. In particular,
the SM prediction suffers from the uncertainties in  the determination 
of the \rm{CKM} element $|V_{ub}|$ and of the decay constant $f_{B}$. 
Concerning $|V_{ub}|$, we use the current \rm{HFAG} average \cite{hfag}
(from combined exclusive and inclusive semileptonic $B$ decays),
while for $f_{B}$ we use the lattice result of~\cite{Latticefb}.
An alternative way to reduce the theoretical error associated to 
${\rm BR}(B\to\tau\nu)_{\rm SM}$ is to consider the
ratio ${\rm BR}(B\to\tau\nu)/\Delta M_{B_d}$, where  $f_{B}$ drops out 
and  $|V_{ub}|$ is replaced 
by $|V_{ub}/V_{td}|=\sin\beta/\sin\gamma$~\cite{Isidori:2006pk,Isidori:2007jw}.
However, as the experimental error on ${\rm BR}(B\to\tau\nu)$ is the 
dominant uncertainty, this alternative way does not lead 
to a significant reduction in the error. Moreover, it complicates the analysis 
since  $\rm{BR}(B\to\tau\nu)$ and $\Delta M_{B_d}$ are affected by 
independent SUSY contributions. For these reasons, we treat the two constraints
separately. More precisely, we treat separately  ${\rm BR}(B\to\tau\nu)$, 
$\Delta M_{B_s}$, and the ratio  $\Delta M_{B_d}/\Delta M_{B_s}$. 
Similarly to \bsg, all flavour-physics constraints apart
from BR($B_{s,d} \to \mu^+\mu^-)$ are implemented 
normalising the observables to the corresponding SM values. 

The direct experimental limit on the Higgs-boson mass in the 
SM obtained at LEP~\cite{Barate:2003sz}
is $\mh > 114.4~\textrm{GeV}$ at the 95\% C.L.
The corresponding bound within the MSSM could in principle be substantially
lower, due to a reduced ZZh coupling or due to different, more complicated
decay modes of the Higgs bosons~\cite{Schael:2006cr}. However, it has been 
shown~\cite{Ellis:2001qv,Ambrosanio:2001xb} that these mechanisms cannot be
realised within the CMSSM, and hence the
experimental lower bound of 114.4~GeV can be applied~\footnote{Following 
Ref.~\cite{Ellis:2007aa}, for simplicity we use this bound also in our
NUHM1 analysis. As discussed below, the best-fit NUHM1 point we find
yields $\mh$ well above this bound.}.
For our fit we use the full likelihood information of the exclusion
bound, given by the $CL_s(\mh)$~value, which is convoluted with a theory error
on the evaluation of $\mh$ of 3~GeV~\cite{Degrassi:2002fi}, according to
the detailed prescription found in~\cite{Ellis:2004tc}.

The numerical evaluation has been performed with the {\tt MasterCode}
that consistently combines the codes responsible for RGE running, for
which we use {\tt SoftSUSY}~\cite{Allanach:2001kg}, and the various
low-energy observables. 
At the electroweak scale we have included the following codes: 
{\tt  FeynHiggs}~\cite{Degrassi:2002fi,Heinemeyer:1998np,Heinemeyer:1998yj,Frank:2006yh} 
for the evaluation of the Higgs masses and $a_\mu^{\rm SUSY}$;
a code based on~\cite{Isidori:2006pk,Isidori:2007jw} and 
{\tt SuperIso}~\cite{Mahmoudi:2008tp} 
for the flavour observables;
a code based on~\cite{Heinemeyer:2006px,Heinemeyer:2007bw} 
for the electroweak precision observables;
{\tt MicrOMEGAs}~\cite{Belanger:2006is,Belanger:2001fz,Belanger:2004yn} and
{\tt DarkSUSY}~\cite{Gondolo:2005we,Gondolo:2004sc} 
for the observables related to dark matter.
We made extensive use of the SUSY Les Houches Accord~\cite{Skands:2003cj}
in the combination of the various codes within the {\tt MasterCode}.

The CMSSM parameter space has been sampled using the MCMC
technique. We treat $m_{1/2}$, $m_0$, $A_0$ and $\tan\beta$ as
free parameters, and the Higgs mixing parameter $\mu$ and the
pseudoscalar Higgs mass $m_A$ as dependent parameters determined by the
electroweak vacuum conditions.

A global $\chi^2$ function is defined, which combines all calculations 
with experimental constraints:
\begin{align}
\chi^2 &= \sum^N_i \frac{(C_i - P_i)^2}{\sigma(C_i)^2 + \sigma(P_i)^2}
       + \sum_i \frac{(f^{\rm obs}_{{\rm SM}_i} 
                  - f^{\rm fit}_{{\rm SM}_i})^2}{\sigma(f_{{\rm SM}_i})^2} 
\end{align}
Here $N$ is the number of observables studied, $C_i$ represents an 
experimentally measured value (constraint) and each $P_i$ defines a CMSSM
parameter-dependent prediction for the corresponding constraint.  The three
SM parameters 
$f_{\rm SM} = \{\Delta\alpha_{\rm had}, m_t, m_Z\}$
are included as fit parameters and constrained to be within their current
experimental resolution $\sigma(f_{\rm SM})$.

As indicated in Section \ref{sec:intro}, the sensitivity of the global fit 
to different constraint scenarios is studied below by removing the
$\Omega_{\rm CDM}$ constraint or rescaling the \gmt\ and other
experimental uncertainties.  Since each 
new scenario represents a new $\chi^2$ function which must be 
minimized, multiple re-samplings of the full multi-dimensional parameter space 
are, in principle, required to determine the most probable fit regions for 
each scenario and would be computationally too expensive.   

To avoid this difficulty, we analyze the effect of removing the
$\Omega_{\rm CDM}$ constraint 
by exploiting the fact that independent $\chi^2$ functions  
are additive and result in a well-defined $\chi^2$ probability.  Hence, a 
``loose'' $\chi^2$ function, $\chi^2_{\rm loose}$, is defined in which the 
term representing the $\Omega_{\rm CDM}$ constraint is removed from the
original $\chi^2$.   The $\chi^2_{\rm loose}$ function represents the
likelihood that a particular set of model parameter values is compatible with
a sub-set of the experimental data constraints, without any experimental
knowledge of $\Omega_{\rm CDM}$.

An exhaustive, and computationally expensive, 25 million point pre-sampling of 
the $\chi^2_{\rm loose}$ function in the full multi-dimensional model parameter 
space is then performed using an MCMC.   The result of 
this pre-sampling identifies fit regions which are generally excluded by the 
considered sub-set of experimental data.  Any regions excluded by the less 
constrained fit will also be excluded with the inclusion of additional 
experimental constraints and, in particular, with different scenarios for the
$\Omega_{\rm CDM}$ constraint.  Hence, without loss of generality, this 
pre-sampling procedure reduces the hyper-volume of parameter space which 
needs to be searched multiple times over in the context of different
constraint scenarios to a computationally manageable level.   

Constraint terms representing the different $\Omega_{\rm CDM}$ scenarios
are then re-instated to form different 
$\chi^2 =  \chi^2_{\rm loose} + \chi^2_{\rm scenario}$ functions, 
one for each scenario studied.  The precise values of the most probable fit 
parameters are determined via a full MINUIT~\cite{James:1975dr} minimization
of the $\chi^2$ for  
each different scenario, but are performed only within the general
parameter space regions not already excluded from the pre-sampling of the 
$\chi^2_{\rm loose}$ function.  An MCMC final sampling 
is subsequently used to determine the 68\% and 95\% likelihood contours for 
each scenario constraint studied. 

Additionally, later on we vary the uncertainties of $\Omega_{\rm CDM}$ and
other constraints using similar techniques. This allows us to study and compare
the effects of such variations on the $\chi^2$~fit and the most
probable parameters.

\section{Results}
\label{sec:results}

\subsection{Predictions for LHC discoveries}

\begin{figure*}[htb!]
\begin{center}
\begin{picture}(300,400)
  \put(  0,   220){ \resizebox{10.5cm}{!}{\includegraphics{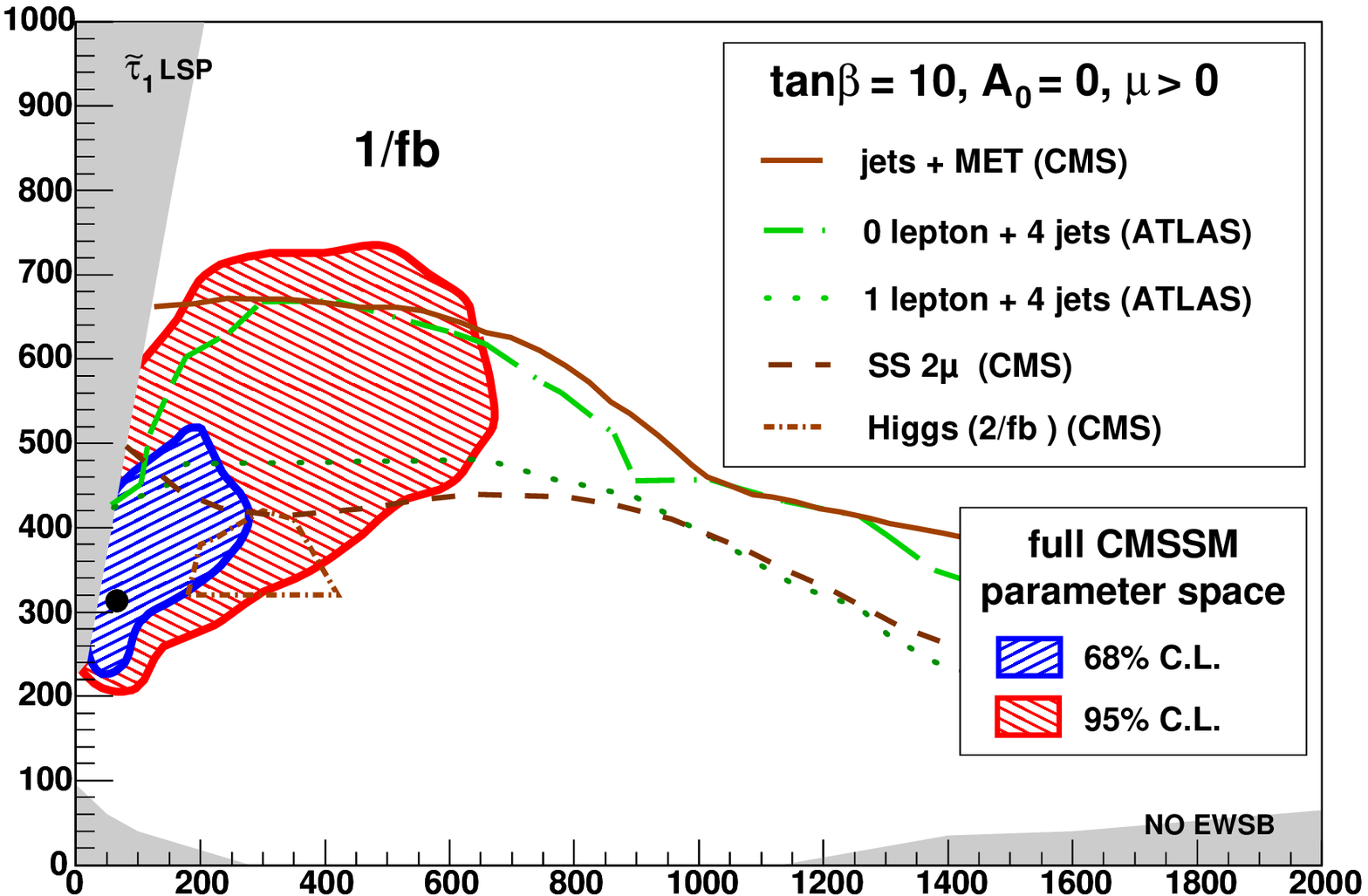}}}
  \put(250,   205){$m_0$ [GeV]}
  \put( -5,   360){\begin{rotate}{90}$m_{1/2}$ [GeV]\end{rotate}}
  \put(  0,     0){ \resizebox{10.5cm}{!}{\includegraphics{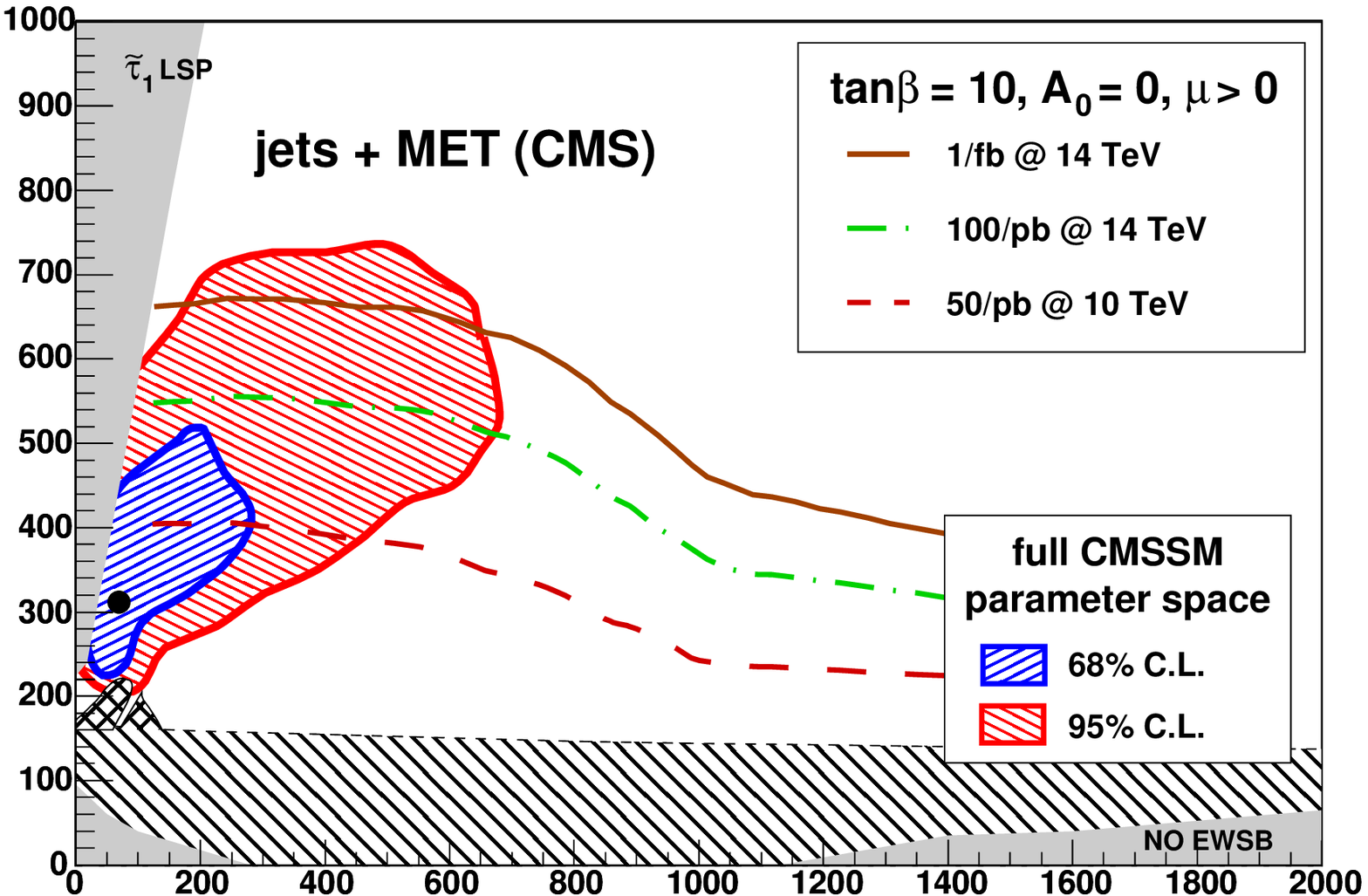}}  }
  \put(250,   -15){$m_0$ [GeV]}
  \put( -5,   140){\begin{rotate}{90}$m_{1/2}$ [GeV]\end{rotate}}
\end{picture}
\end{center}
\vspace{-0.5em}
\caption {The $(m_0, m_{1/2})$ plane in the CMSSM for $\tan\beta = 10$
  and $A_0 = 0$. The dark shaded area at low $m_0$ and high $m_{1/2}$ is
  excluded due 
  to a scalar tau LSP, the light shaded areas at low $m_{1/2}$ do not
  exhibit electroweak symmetry breaking. The nearly horizontal line at
  $m_{1/2} \approx 160$~GeV in the lower panel 
  has $m_{\tilde \chi_1^\pm} = 103$~GeV, and the area
  below is excluded by LEP searches. Just above this contour at low $m_0$
  in the lower panel is the region that is
  excluded by trilepton searches at the Tevatron.
  Shown in both plots are the best-fit point, indicated by a filled
  circle, and the 
  68 (95)\%~C.L.\ contours from our fit as dark grey/blue (light
  grey/red) overlays, scanned over all $\tan\beta$ and $A_0$ values.
  Upper plot: Some $5\,\sigma$ discovery contours at ATLAS and CMS with
  1~fb$^{-1}$ at 14~TeV, and the contour for the $5\,\sigma$ discovery
  of the Higgs boson in sparticle decays with 2~fb$^{-1}$ at 14~TeV in CMS.
  Lower plot: The $5\,\sigma$ discovery contours for jet + missing $E_T$ events 
  at CMS with 1~fb$^{-1}$ at 14~TeV, 100~pb$^{-1}$ at 14~TeV and
  50~pb$^{-1}$ at 10~TeV centre-of-mass energy.
} 
\label{fig:contours}
\end{figure*}

In Fig.~\ref{fig:contours} we display the best-fit value and the 
68\% and 95\% likelihood contours 
for the CMSSM $(m_0, m_{1/2})$ plane,
obtained as described in Sect.~\ref{sec:mpfit} from a fit taking into
account all experimental constraints listed in Tab.~\ref{tab:constraints}
as well as the constraints from the additional electroweak observables listed 
in~\cite{Buchmueller:2007zk}.
We also show in the upper panel of Fig.~\ref{fig:contours} various LHC
sparticle discovery contours for 1~fb$^{-1}$ of good-quality data
in a single experiment
at a centre-of-mass energy of 14~TeV. 
The ATLAS and CMS collaborations have each published
5-$\sigma$ discovery contours in the CMSSM $(m_{1/2}, m_0)$ plane for
$A_0 = 0$ and $\tan \beta = 10$~\cite{atlastdr,cmstdr,Blaising}. 
Their contours are generally very similar, and the solid brown contour 
displayed 
is that published by CMS for the most sensitive jets + missing $E_T$ search. 
This contour is insensitive to $A_0$, which affects primarily the
third-generation sparticle masses, since the main discovery channels
involve gluinos and first-generation squarks. The discovery contours are
also not very sensitive to $\tan \beta$, since the gluino mass is
insensitive to this variable, and the first-generation squark masses are
also not very sensitive to $\tan \beta$. Therefore, it is a reasonable first
approximation to compare our  68\% and 95\% likelihood contours
directly with the discovery contours given for $A_0 = 0$ and 
$\tan\beta = 10$ fixed, particularly since the best fit has a similar
value of $\tan \beta$. 

\begin{figure*}[htb!]
\begin{picture}(450,210)
  \put(  5,  10){ \resizebox{7.7cm}{!}{\includegraphics{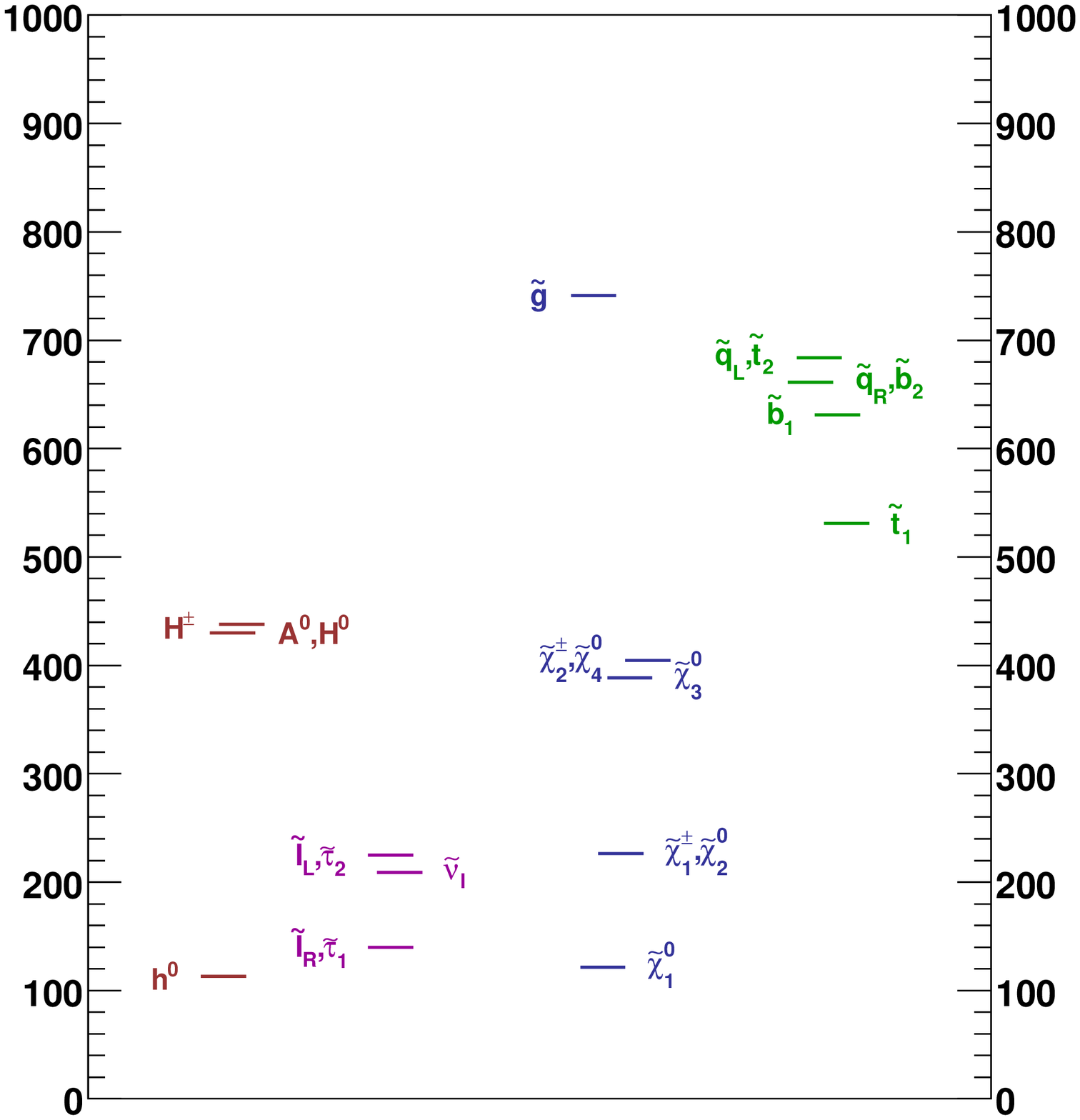}}  }
  \put( 10, 185){\begin{rotate}{90}$m$ [GeV]\end{rotate}}
  \put(140, 200){CMSSM}
  \put(235,  10){ \resizebox{7.7cm}{!}{\includegraphics{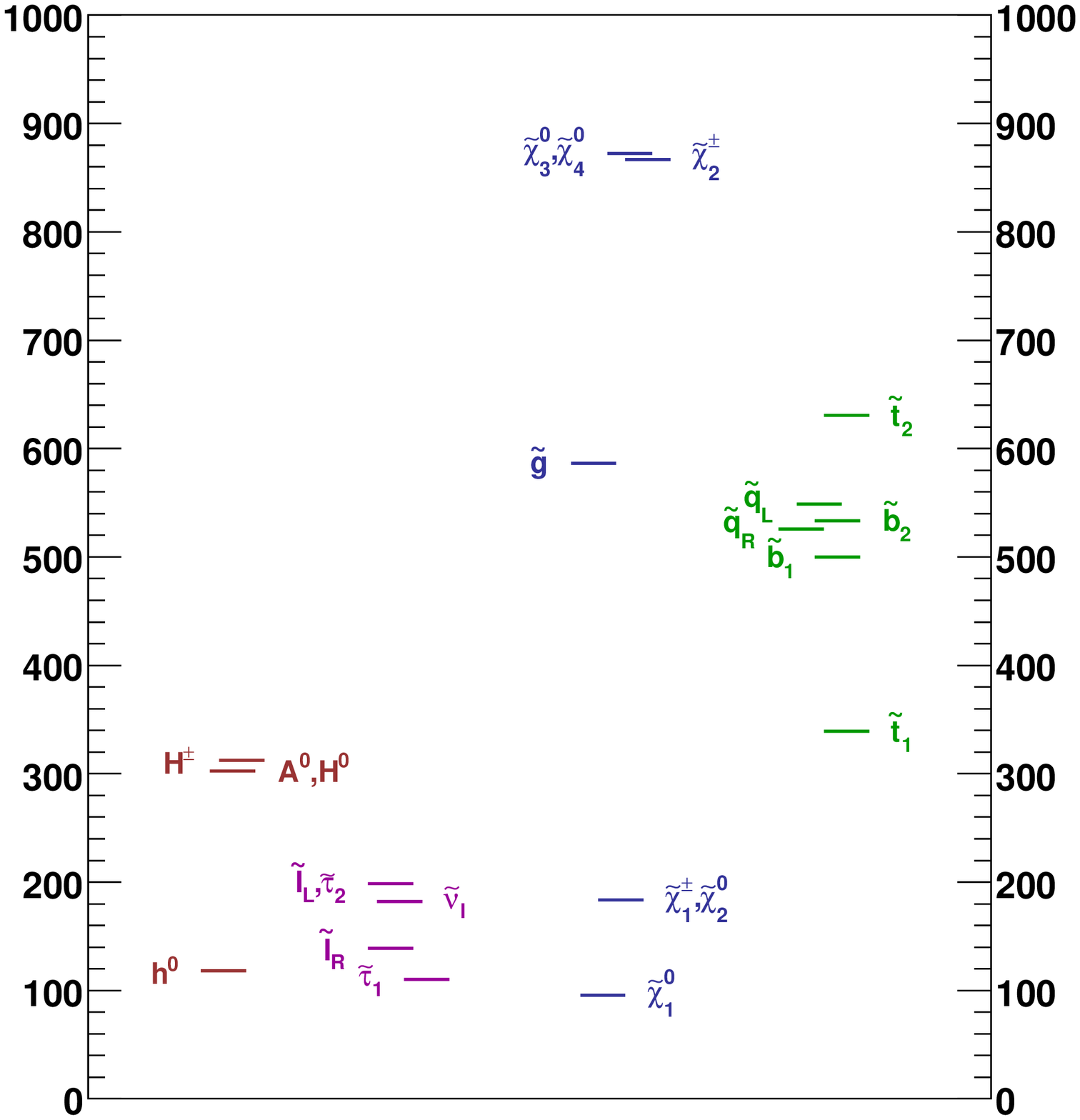}}  }
  \put(240, 185){\begin{rotate}{90}$m$ [GeV]\end{rotate}}
  \put(370, 200){NUHM1}
\end{picture}
\vspace{-2em}
\caption {The spectra at the best-fit points: left --- in the CMSSM with 
$m_0 = 60$~GeV, $m_{1/2} = 310$~GeV, $A_0 = 240$~GeV, $\tan\beta = 11$, and
right --- in the NUHM1 with $m_0 = 100$~GeV, $m_{1/2} = 240$~GeV, 
$A_0 = -930$~GeV, $\tan \beta = 7$, $m_H^2 = - 6.9 \times 10^5$~GeV$^2$
and $\mu = 870$~GeV.}
\label{fig:spectra}
\end{figure*}

The parameters of the best-fit CMSSM point are
$m_{1/2} = 310$~GeV, $m_0 = 60$~GeV,   $A_0 = 240$~GeV, $\tan \beta = 11$
and $\mu = 380$~GeV~\footnote{Here and later, we quote CMSSM and NUHM1
input mass parameters with 10~GeV accuracy.}, 
yielding the overall $\chi^2/{\rm N_{dof}} = 20.4/19$ (37.3\% probability) 
and $\mh = 113.2$~GeV~\footnote{The CMSSM fit quality has improved relative 
to~\cite{Buchmueller:2007zk} primarily because of the new value of $m_t$
and the inclusion of more observables, that are generally highly
consistent with the CMSSM.}. 
The overall value of the $\chi^2$ at the minimum is somewhat pushed up by
the value of $\mh$, which is uncomfortably low. However, it is
acceptable within the higher-order calculational uncertainties expected
in the {\tt FeynHiggs} code that we use here, 
$\delta\mh^{\rm theo} \approx$~3~GeV~\cite{Degrassi:2002fi}. 
As we discuss below, this slight tension is removed in the NUHM1 model,
which correspondingly has a somewhat lower overall $\chi^2$ 
(yielding a similar fit probability for the two models).
The spectrum at the best-fit CMSSM point is shown in the left panel of 
Fig.~\ref{fig:spectra}. 
It is interesting to note that the best-fit CMSSM point and the
corresponding spectrum are quite similar to the well-known SPS1a
benchmark point~\cite{sps}, whose phenomenology at future colliders has
been studied in considerable detail (see, e.g.,
\cite{bench1,Battaglia:2003ab,lhclc}).

Comparing the 95\% likelihood contour provided by the multi-parameter fit
with the 1~fb$^{-1}$ LHC discovery contour, we see that the former is almost
entirely contained within the latter, implying that, if the CMSSM were
correct, the LHC would be almost  `guaranteed', with 95\% confidence, to
discover SUSY with 1~fb$^{-1}$ of good-quality data at 14~TeV.
We also display in the upper panel of
Fig.~\ref{fig:contours} contours representing the 5~$\sigma$ 
discovery reach with 1~fb$^{-1}$ at 14~TeV for 4-jet events with and
without a charged lepton, for same-sign dileptons~\cite{cmstdr}, denoted
by SS, and (with 2~fb$^{-1}$) for the lightest MSSM 
Higgs boson produced in cascade decays of sparticles~\cite{cmstdr},
denoted by $h$. We see that the same-sign dilepton discovery region
largely covers the 68\% likelihood region of the CMSSM $(m_0, m_{1/2})$
plane. Thus, this signature  
could serve as a clean signal capable of confirming the supersymmetric
interpretation of any jets + missing $E_T$ signal observed in initial
LHC running. On the other hand, the region where the lightest Higgs
  boson could be discovered in cascade decays of squarks with
  2~fb$^{-1}$ at 14~TeV lies largely between the 95\% and 68\%
  C.L.\ contours.

We have used {\tt PROSPINO2}~\cite{Prospino} to estimate the variation
of the discovery reach of the LHC jets + missing $E_T$
search as a function of the integrated luminosity and the
centre-of-mass energy. We display in the lower panel of
Fig.~\ref{fig:contours} (green) dot-dashed and (red) dashed contours
representing, 
respectively, the discovery reaches expected with 100~pb$^{-1}$ at
14~TeV  and 50~pb$^{-1}$ at 10~TeV. 
We see that the 68\% likelihood contour is
well covered by the 14~TeV/100~pb$^{-1}$ discovery reach, and even
the 10~TeV/50~pb$^{-1}$ reach would be sufficient to discover SUSY at
the best-fit point, indicated by a filled circle in Fig.~\ref{fig:contours}. 
The lower panel of Fig.~\ref{fig:contours} also displays the regions
of the CMSSM $(m_0, m_{1/2})$ plane that are
excluded by chargino searches at LEP and by sparticle searches at the 
Tevatron~\cite{pdg,d0:2007ww,cdfsusy}.
The region excluded by the LEP Higgs search is sensitive to $\tan \beta$
and $A_0$, is subject to theoretical uncertainties, and, moreover, the
experimental Higgs 
likelihood function is not a simple step function. Hence, it is not shown in
Fig.~\ref{fig:contours}~\footnote{However, 
for orientation, we note that if $m_0 = 0$, $\tan \beta = 10$ and $A_0 = 0$
the evaluation with 
{\tt FeynHiggs} yields a nominal value of $\mh = 114.4$~GeV for
$m_{1/2} = $307~GeV.}.

\subsection{Sensitivity to experimental constraints}
\label{sec:sensitivity}

The above analysis assumed the default implementations of the experimental, 
phenomenological and cosmological constraints discussed in the previous
Section. We now discuss the possible effects of relaxing (or
strengthening) some of the key constraints, starting with the relic cold
dark matter density, $\Omega_{\rm CDM}$. 

It is well-known that this constraint essentially reduces the
dimensionality of the MSSM parameter space by one unit, fixing one
combination of the parameters with an accuracy of a few~\%. For example,
in the CMSSM for any pair of fixed values of $A_0$ and $\tan \beta$, the
$\Omega_{\rm CDM}$ constraint largely determines $m_0$ as a function of
$m_{1/2}$, except for a discrete ambiguity associated with the
coannihilation strip, the focus-point strip and the rapid-annihilation
funnel that appears at large $\tan \beta$. Therefore, one might expect
that dropping the  $\Omega_{\rm CDM}$ constraint would have a strong
effect on the preferred region of the CMSSM $(m_{1/2}, m_0)$ plane shown
in Fig.~\ref{fig:contours}. 

There are various possible reasons why one might consider dropping the
dark matter 
constraint. Perhaps the neutralino is not the LSP? Perhaps $R$-parity is
not quite conserved? Perhaps the early thermal history of the Universe
differed from that usually assumed when calculating the relic LSP
density? Perhaps Nature is described by some generalization of the CMSSM
such as a model with non-universal SUSY-breaking contributions to the
Higgs scalar masses (NUHM), in which case values of $m_0$ very different
from those in the CMSSM might be permitted?

\begin{figure*}[htb!]
\begin{picture}(440,412) %
  \put(-10, 220){ \resizebox{9cm}{!}{\includegraphics{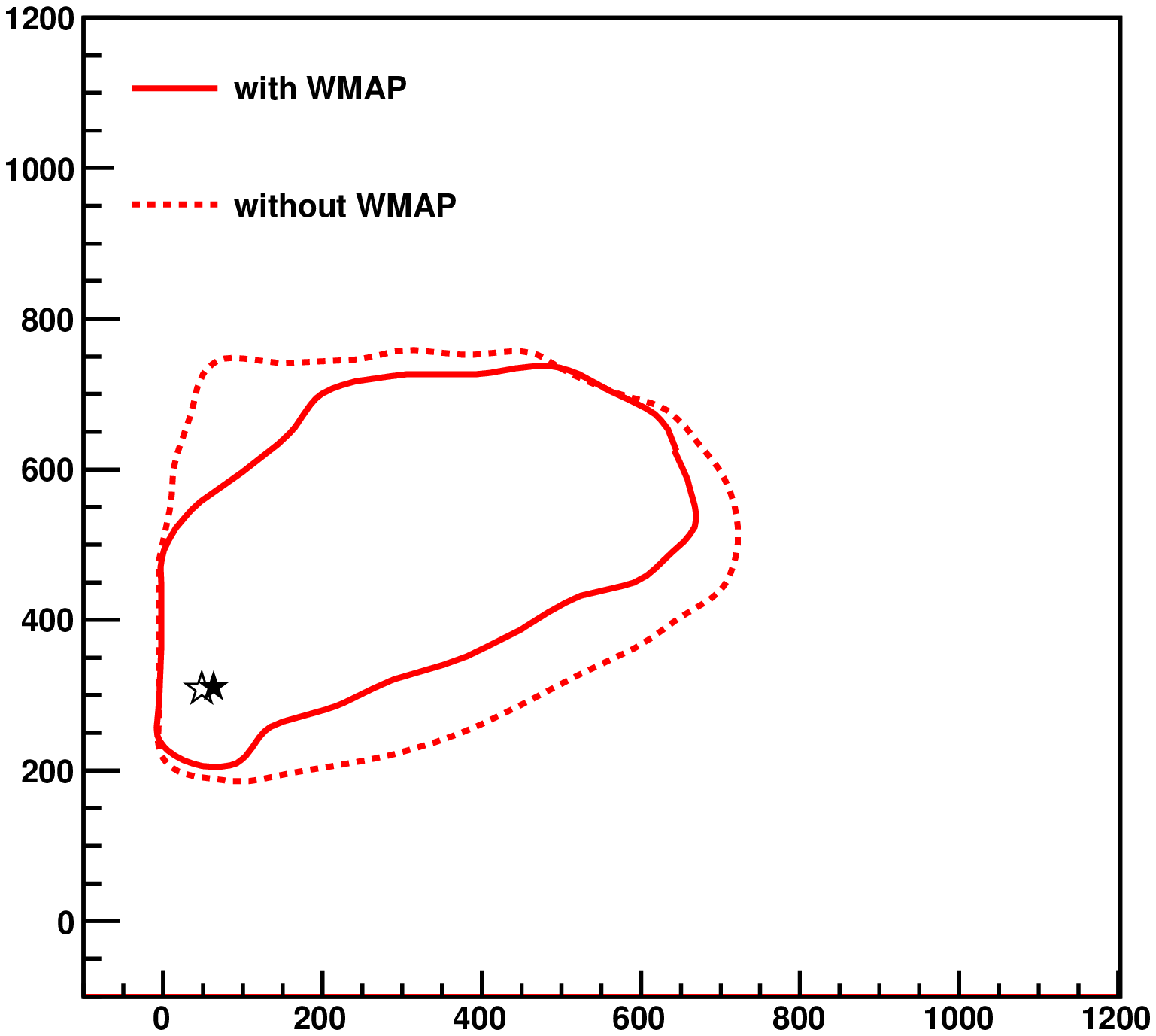}}  }   
  \put(220, 220){ \resizebox{9cm}{!}{\includegraphics{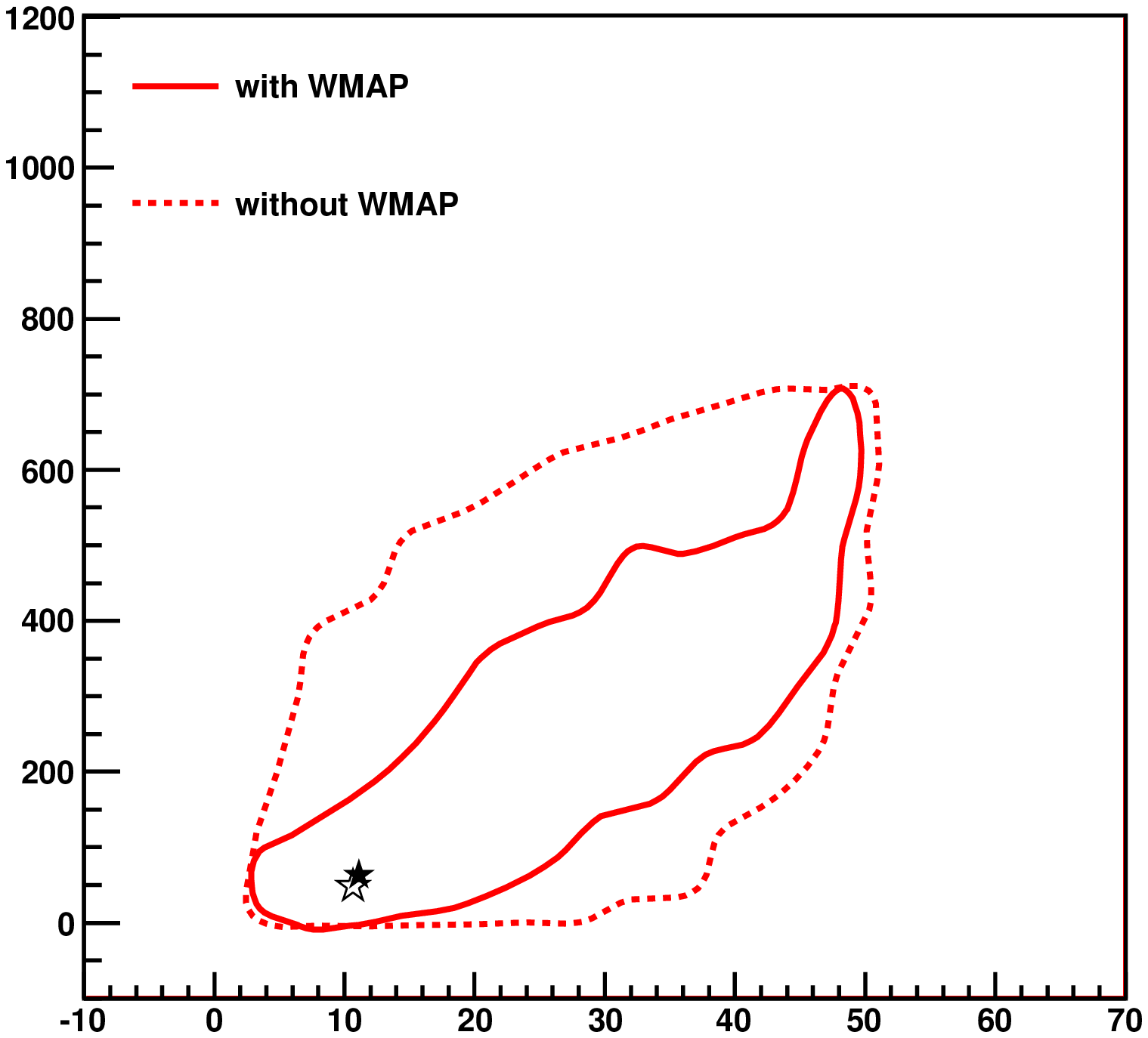}}  }  
  \put(-10,   0){ \resizebox{9cm}{!}{\includegraphics{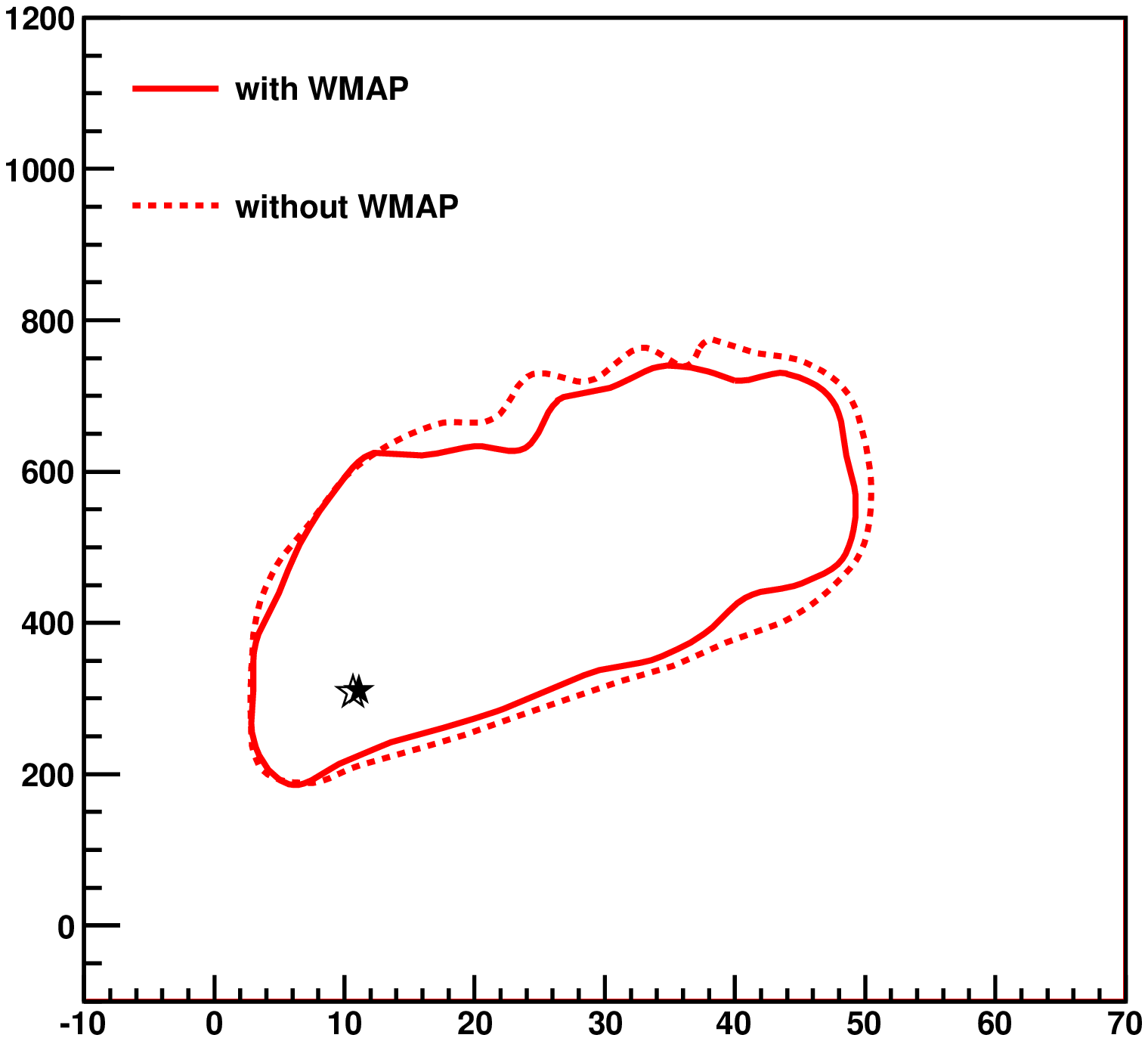}}  }  
  \put(220,   0){ \resizebox{9cm}{!}{\includegraphics{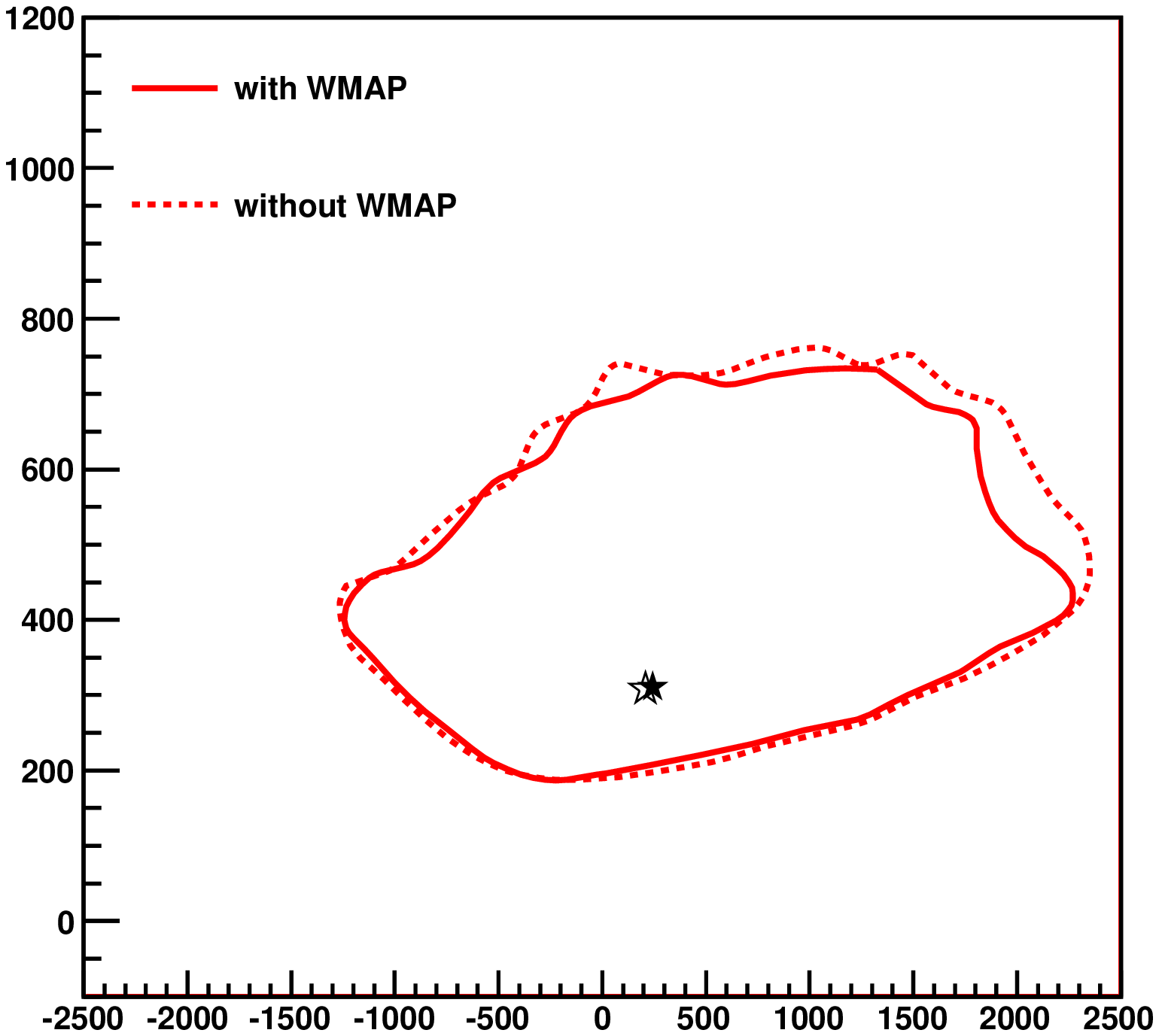}}  }   
  \put( 10, 360){\begin{rotate}{90}$m_{1/2}$ [GeV]\end{rotate}}
  \put(170, 220){$m_0$ [GeV]}
  \put(240, 360){\begin{rotate}{90}$m_0$ [GeV]\end{rotate}}
  \put(415, 220){$\tan\beta$}
  \put( 10, 148){\begin{rotate}{90}$m_{1/2}$ [GeV]\end{rotate}}
  \put(185,   0){$\tan\beta$}
  \put(240, 140){\begin{rotate}{90}$m_{1/2}$ [GeV]\end{rotate}}
  \put(400,   0){$A_0$ [GeV]}
\end{picture}
\vspace{-1em}
\caption{Variation of the 95\%~C.L.\ allowed regions in the MSSM
  parameter space including (solid) or excluding (dotted) the WMAP constraint. 
  The plots show the 
$(m_0, m_{1/2})$~plane (upper left), $(\tan\beta, m_0)$~plane (upper
right), $(\tan\beta, m_{1/2})$~plane (lower left) and the 
$(A_0, m_{1/2})$~plane (lower right plot). In each panel, we mark the 
best-fit points
found both with and without the WMAP constraint by a filled and open
star, respectively.
}
\label{fig:WMAP}
\end{figure*}

We show in Fig.~\ref{fig:WMAP} the effect of dropping the
$\Omega_{\rm CDM}$ constraint. This is significant in the upper panels, 
which display the $(m_0, m_{1/2})$ and $(\tan \beta, m_0)$ planes,
but  is not so important in the
$(\tan\beta,m_{1/2})$ and $(A_0, m_{1/2})$~planes shown in the two lower
panels of Fig.~\ref{fig:WMAP}.
These behaviours can be understood by
recalling the behaviour of the WMAP 
coannihilation strips in the CMSSM $(m_{1/2}, m_0)$
planes for different values of $\tan \beta$. For example, the
value of $m_0$ favoured by $\Omega_{\rm CDM}$ for any given values of 
$m_{1/2}$ and $A_0$ increases as the value
of $\tan \beta$ increases, foliating the $(m_0, m_{1/2})$ plane. Thus, 
for any given value of $m_{1/2}$ and $A_0$, a
large range of values of $m_0$ can be attained for a suitable choice of
$\tan \beta$, even if one does impose the  $\Omega_{\rm CDM}$
constraint. Concerning the range of $m_{1/2}$, this is bounded above by
\gmt, and, for any given value of $\tan \beta$, the allowed range
actually decreases for the larger values of $m_0$ allowed if the
$\Omega_{\rm CDM}$ constraint is dropped. Thus, dropping
the $\Omega_{\rm CDM}$ constraint has little overall effect on the ranges of
$m_{1/2}$, $m_0$ and $A_0$. The primary effect is to enforce a correlation
between $\tan \beta$ and $m_0$, as seen in the upper right panel of
Fig.~\ref{fig:WMAP}. The range of $m_0$ decreases at any fixed value of 
$\tan \beta$ when the $\Omega_{\rm CDM}$ constraint is
imposed, because of the
narrowness of the WMAP strip for any fixed value of $\tan \beta$.

Thus, we find that the fit results obtained in the parameter planes of
$m_0$, $m_{1/2}$, $A_0$ and $\tan\beta$ displayed in
Fig.~\ref{fig:contours} are rather robust with respect to imposing / dropping 
the $\Omega_{\rm CDM}$ constraint. On the other hand, as already noted,
the $\Omega_{\rm CDM}$ constraint does reduce the dimensionality of the
parameter space by essentially one
unit. This can also be seen from the fact that without imposing the 
$\Omega_{\rm CDM}$ constraint the preferred parameter region obtained
from the fit to the EWPO and BPO still yields a wide range of possible 
values of $\Omega_{\rm CDM} h^2$. Specifically we find that the 68\% C.L.
region of the $(m_0, m_{1/2})$ plane shown in Fig.~\ref{fig:contours} yields
$\Omega_{\rm CDM} h^2 < 0.9$, while considerably larger values
of $\Omega_{\rm CDM} h^2$ are allowed at the 95\% C.L. Eventually, SUSY
particle mass measurements
at the LHC (see the discussion of Fig.~\ref{fig:dileptons} below)
may enable this estimate of $\Omega_{\rm CDM} h^2$ to be
refined considerably (see, 
e.g., \cite{Battaglia:2003ab,Nojiri:2005ph,Baltz:2006fm}).

\begin{figure*}[htb!]
\vspace{1em}
\begin{picture}(450,155)
  \put(5,   0){ \resizebox{8.1cm}{!}{\includegraphics{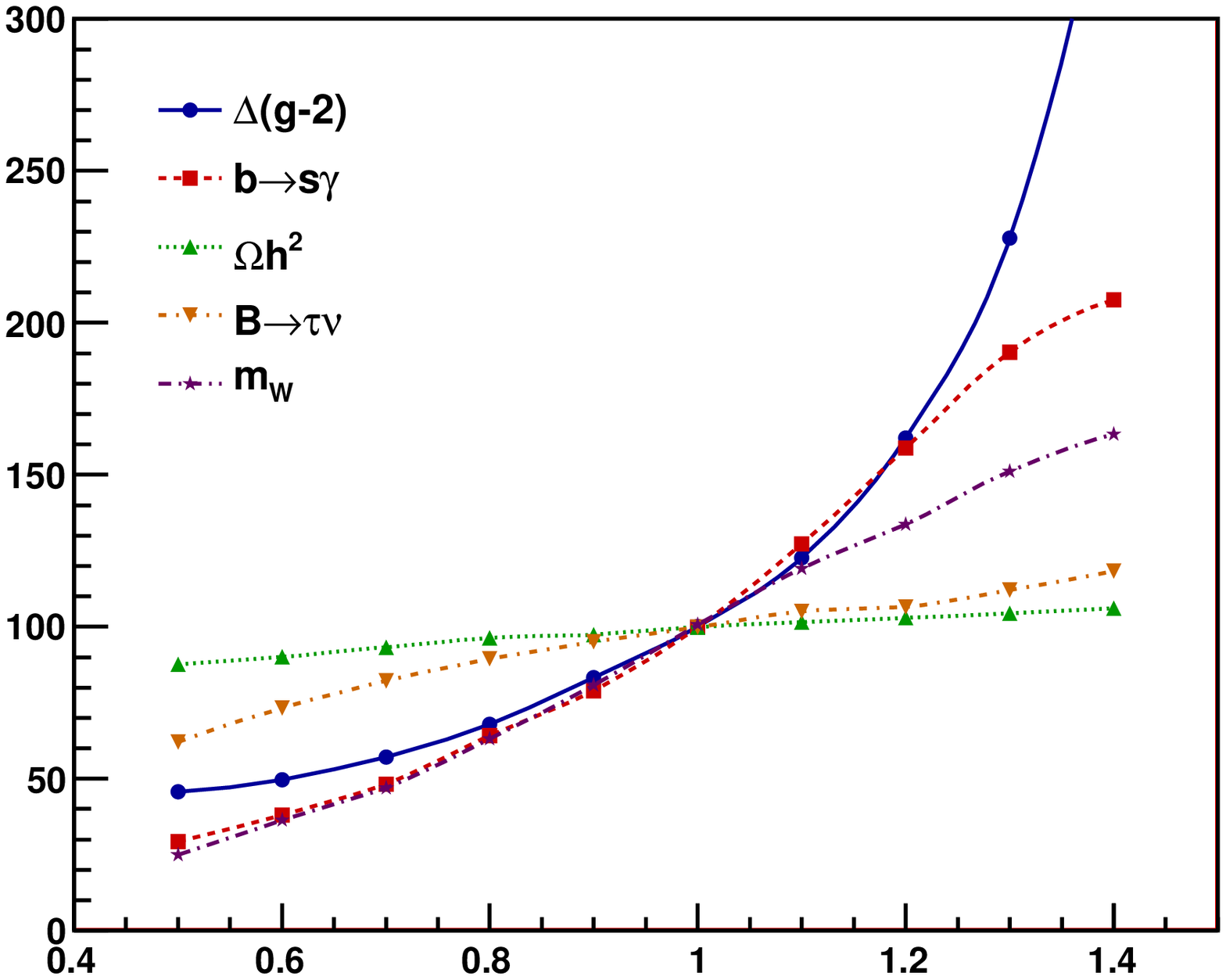}}  }
  \put(235, 0){ \resizebox{8.1cm}{!}{\includegraphics{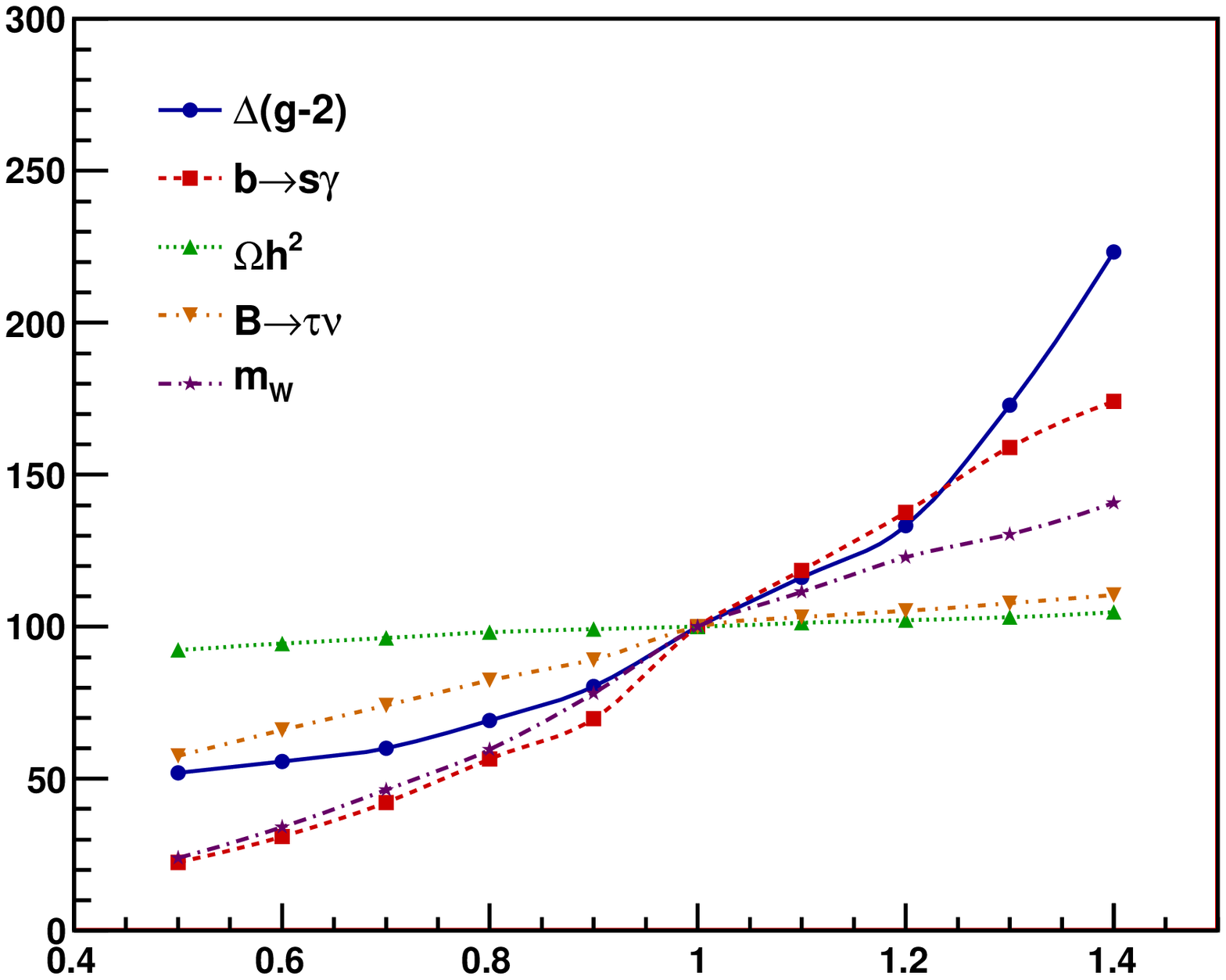}}  }
  \put(10,43){\begin{rotate}{90}Relative Change in Area [\%]\end{rotate}}
  \put(77,0){Relative Change in Uncertainty}
  \put(120,138){$(m_0,m_{1/2})$}
  \put(240,43){\begin{rotate}{90}Relative Change in Area [\%]\end{rotate}}
  \put(307,0){Relative Change in Uncertainty}
  \put(350,138){$(m_0,\tan\beta)$}
\end{picture}
\caption {Relative sizes of the 95\%~C.L.\ areas in the $(m_0, m_{1/2}$)
  plane (left) and in the $(m_0, \tan\beta)$ plane (right) as a function
  of the hypothetical errors of \gmt, \bsg, 
  $\Omega_{\rm CDM}h^2$, \btn, \MW. The error
  scaling is relative to the current combined theory and experimental
  error. 
}
\label{fig:drill}
\end{figure*}

Drilling down into the dependences of our results on uncertainties in
the experimental and phenomenological constraints, we display in
Fig.~\ref{fig:drill} the results of studies of their sensitivities
to some key observables. The observables tracked are \gmt, 
\bsg, $\Omega_{\rm CDM} h^2$, \btn\ and \MW. The
left panel shows the  percentage variation in the preferred region of
the $(m_0, m_{1/2})$ plane as the assumed errors in these quantities are
rescaled, assuming that the future experimental central values agree
with the current ones. 
The right panel shows the same for the area in the $(m_0,
\tan \beta)$ plane. Larger errors could arise if we have underestimated
the relevant systematic errors, and smaller errors could result from
future improvements of the experimental errors and/or the theoretical
predictions. As could be expected from
the discussion in the previous
paragraph, the preferred areas vary very little with the error in
$\Omega_{\rm CDM}$, and the areas are also relatively insensitive to
that in \btn. However, there are greater sensitivities
to \bsg, \MW\ and (particularly) \gmt .

The theoretical error in \MW\ is much smaller than the current
experimental error. It is encouraging that reducing the experimental
error, as should be possible with future Tevatron and LHC data, could
have substantial effects on the preferred areas in the parameter planes.
A reduction in the error by a factor two could reduce the areas by
factors of about five, if the present central value (which disagrees with
the SM by about one $\sigma$) is maintained. 

The same would be true for a reduction in the error in \bsg, but here reducing 
the theoretical error would also be necessary. This would require, in
particular, a better understanding of the uncertainties in higher-order
and non-perturbative
QCD corrections. Indeed, a very conservative approach to the combination
of the current theoretical and experimental errors in \bsg\ might even
motivate a larger error and hence larger preferred areas than in our
default analysis.  

Fig.~\ref{fig:drill} also shows that varying the error in \gmt\ is
potentially more important, particularly if the present error is
underestimated. This might be the case if, e.g., the weight of
experimental evidence would shift towards using $\tau$ decay data to
estimate the SM hadronic contribution to \gmt, or if the error in
the light-by-light contribution were to be revised
drastically~\footnote{In \cite{Passera:2008jk} it has recently been
claimed  that solving the muon $(g-2)_{\mu}$ anomaly 
by changing the SM prediction of the hadronic
contribution to $(g-2)_{\mu}$ is unlikely in view of a
combined analysis of all electroweak data.}.
The rapid increases in the areas of the preferred regions reflect the fact
that a more relaxed treatment of the \gmt\ error led in the past
to (parts) of the focus-point strip at large $m_0$ being included within
the preferred region, which does not occur in our default analysis. 

\begin{figure*}[htb!]
\begin{picture}(450,186)
  \put(0,   0){ \resizebox{8.80cm}{!}{\includegraphics{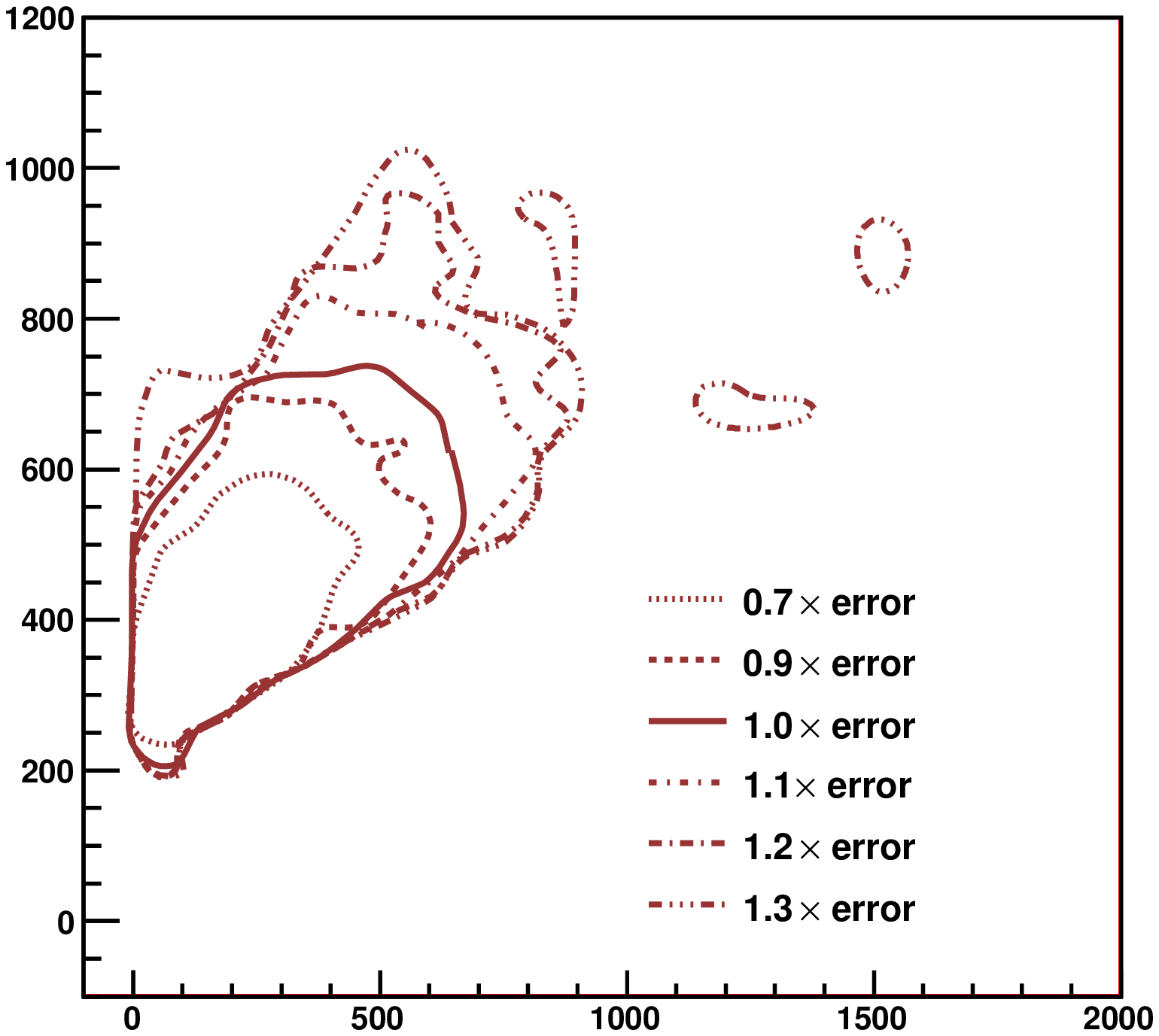}}  }
  \put(225, 0){ \resizebox{8.80cm}{!}{\includegraphics{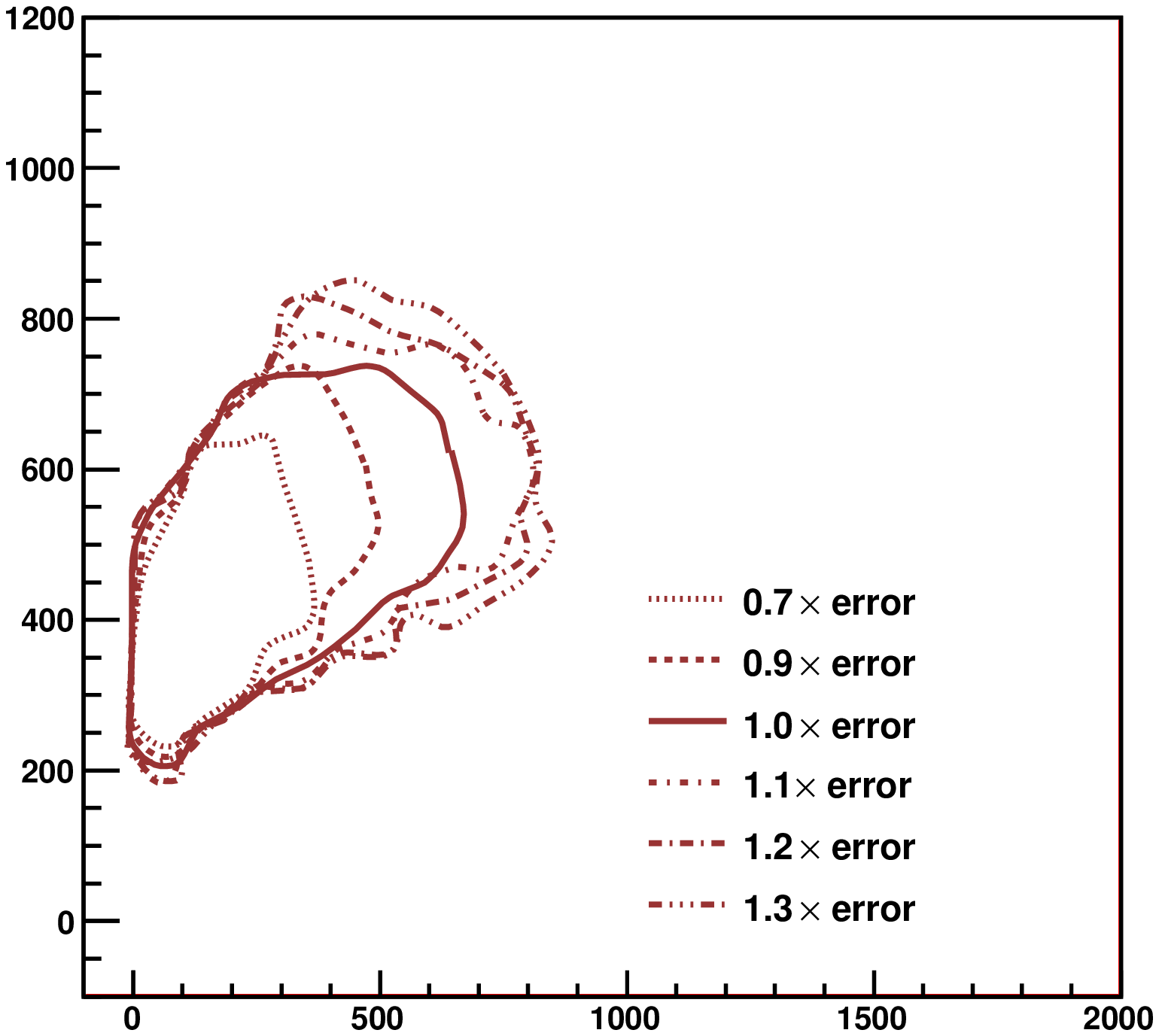}}  }
  \put(20,138){\begin{rotate}{90}$m_{1/2}$ [GeV]\end{rotate}}
  \put(175,5){$m_0$ [GeV]}
  \put(245,138){\begin{rotate}{90}$m_{1/2}$ [GeV]\end{rotate}}
  \put(400,5){$m_0$ [GeV]}
\end{picture}
\vspace{-2em}
\caption {The 95\%~C.L.\ region in the $(m_0, m_{1/2})$~plane for various
  hypothetical values of the $1\,\sigma$~uncertainty (combined theory and
  experimental) of \gmt (left) and \bsg (right). The curves show (from
  outer to inner) the 95\%~C.L.\ regions for 
$\sigma_{\rm hypothetical}/\sigma_{\rm today} = 1.3, 1.2, 1.1, 1.0, 0.9, 0.7$ for 
$\sigma_{\rm today} = 8.8 \times 10^{-10}$ (left) and 
$\sigma_{\rm today} ( {\rm BR}_{\rm b \to s \gamma}^{\rm exp}/
                    {\rm BR}_{\rm b \to s \gamma}^{\rm SM}) = 0.12$ 
(right), respectively, see Tab.~\ref{tab:constraints}. 
}
\label{fig:g-2}
\end{figure*}

In order to explore the sensitivity to the \gmt\ error in more detail, we show 
in the left panel of Fig.~\ref{fig:g-2} the effect in the CMSSM $(m_0,
m_{1/2})$ plane of varying this error, while assuming the
same central value. Going from the outer
to the inner contours we have assumed 
$\sigma_{\rm hypothetical}/\sigma_{\rm today} = 1.3, 1.2, 1.1, 1.0, 0.9, 0.7$
with $\sigma_{\rm today} = 8.8 \times 10^{-10}$, see
Tab.~\ref{tab:constraints}. The partially fuzzy shapes would be smoothed
by higher statistics.
We see that the preferred region expands  rapidly if the
\gmt\ error is increased. Going to an increase by a factor of~1.5
(not shown in the plot) would open up the focus-point region, which is
disfavoured in our analysis.
Conversely, decreasing this error, as
would be possible with an accessible improvement of the previous BNL
\gmt\ experiment~\cite{Hertzog:2007hz}, would enable the preferred
ranges of the CMSSM 
mass parameters to be decreased impressively. Ultimately, this together
with the other EWPO and BPO could
make possible a sensitive test of SUSY at the loop level, if the LHC
does indeed discover sparticles and measure their masses.

In the right panel of Fig.~\ref{fig:g-2}, we make a similar
analysis of the sensitivity to the \bsg\ error.
Going from the outer to the inner contours we have again assumed
$\sigma_{\rm hypothetical}/\sigma_{\rm today} = 1.3, 1.2, 1.1, 1.0, 0.9, 0.7$
with $\sigma_{\rm today}({\rm BR}_{\rm b \to s \gamma}^{\rm exp}/
{\rm BR}_{\rm b \to s \gamma}^{\rm SM}) = 0.12$, see Tab.~\ref{tab:constraints}.
We see from the right panel of Fig.~\ref{fig:g-2} that treating the
errors differently 
could have a significant effect.   Employing a larger error (as done
in~\cite{Feroz:2008wr}, for instance), would not only expand the allowed
regions, but also allow larger $\tan\beta$ values, as \bsg\  
is particularly sensitive to $\tan\beta$.

Though in the present analysis we have focused on the $\mu>0$ solution,
as favoured by the \gmt\ anomaly, we comment briefly here on the 
structure of the $\mu<0$ parameter space. In order to minimize the
discrepancy with the \gmt\ constraint, for $\mu<0$ one would need a
relatively heavy spectrum 
in order to suppress the SUSY effects with the wrong sign.
This would be particularly true for increasing values of $\tan\beta$,
since \gmt\ grows almost linearly with $\tan\beta$.
\bsg\ is also highly sensitive to the sign of the $\mu$ parameter. 
In particular, within the CMSSM the solution with $\mu<0$ unambiguously 
implies that all the dominant SUSY effects to \bsg\ have the same sign
and interfere constructively with the SM amplitude. 
This implies more severe constrains with respect to the $\mu>0$ case,
and again points toward a heavy spectrum. This is not the case for $\mu>0$,
where partial cancellations among SUSY effects in  \bsg\
allow relatively light squarks. 

We have also considered possible improvements in the determination of
the CMSSM parameters that might be obtainable from early LHC measurements. 
Missing $E_T$ measurements with or without single leptons are unlikely to
constrain the model with high precision. On the other hand, in the
parameter region preferred by the fit (with $\tan\beta \approx 10$)
there are good prospects for measuring the 
opposite-sign dilepton edge in 
$\tilde\chi_2 \to \tilde\chi_1 \ell^+ \ell^-$ ($\ell = e, \mu$)
decays with high precision, which is located at
\begin{equation}
(m_{\ell\ell}^2)^{\textrm{edge}} =
\frac{(m^2_{\tilde{\chi}_2^0}-m^2_{\tilde{\ell}_R})(m^2_{\tilde{\ell}_R}-
m^2_{\tilde{\chi}_1^0})}{m^2_{\tilde{\ell}_R}}.
\label{edge}
\end{equation}
Such a measurement would constrain a combination of sparticle masses
and hence the CMSSM parameter space in an interesting way. As an appetizer for
what might be possible, we show in Fig.~\ref{fig:dileptons} the
possible impact of a measurement of the dilepton edge for the CMSSM
best-fit point described in the previous paragraph,
which has $m_{\tilde\chi_1^0} = 121$~GeV, $m_{\tilde\chi_2^0}  = 225$~GeV,
$m_{\tilde l_R} = 139$~GeV, 
yielding an edge at $m_{\ell^+ \ell^-} = 87$~GeV. 
We assume experimental and theoretical errors of 3~GeV each.
We see in Fig.~\ref{fig:dileptons} that the
dilepton edge measurement would reduce the parameter space preferred at
the 68\% C.L.\ to two narrow strips in the $(m_0, m_{1/2})$ plane,
linked into a tilted `vee' shape at the 95\% C.L. The best-fit point in
the right wing of the `vee' has quite different parameter values from
the overall best-fit point in the left wing of the `vee':  
$m_{1/2} = 390$~GeV, $m_0 = 230$~GeV, $A_0 = 1230$~GeV, 
$\tan \beta = 23$, yielding $\chi^2 = 22.7$ and 
$m_{\tilde\chi_1^0} = 155$~GeV, $m_{\tilde\chi_2^0}  = 293$~GeV, 
$m_{\tilde l_R} = 273$~GeV.

\begin{figure*}[htb!]
\begin{center}
\begin{picture}(325,186)
  \put(0,  0){ \resizebox{11cm}{!}{\includegraphics{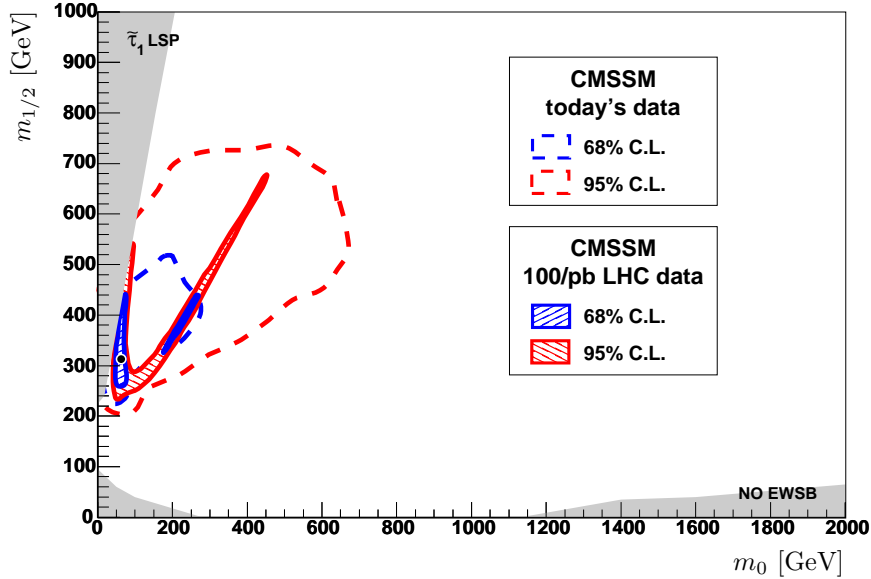}}  }
  \put(-5,149){\begin{rotate}{90}$m_{1/2}$ [GeV]\end{rotate}}
  \put(261,-13){$m_0$ [GeV]}
\end{picture}
\end{center}
\vspace{-0.5em}
\caption{The $(m_0, m_{1/2})$ plane in the CMSSM for $\tan\beta = 10$
  and $A_0 = 0$, showing the improvement in the constraints on $m_0$ and
  $m_{1/2}$ that could be obtained by measuring the opposite-sign
  dilepton edge with   1~fb$^{-1}$ of integrated luminosity at 14~TeV,
  assuming the spectrum of the best-fit 
  point shown in the left panel of Fig.~\protect\ref{fig:spectra},
   and experimental and theoretical errors of 3~GeV each.
  The best-fit point is indicated by a filled circle.
  } 
\label{fig:dileptons}
\end{figure*}

\subsection{Comparison with the NUHM1 case}

The above analysis of the CMSSM is relatively encouraging for the early days
of the LHC, but one might wonder to what extent the conclusions can be extended
to more general incarnations of the MSSM. The full parameter space of
the MSSM has so many 
dimensions that exploring it with the MCMC approach used here would require
prohibitive amounts of CPU time. Accordingly, we discuss briefly here only the
simplest possible generalization of the CMSSM, in which the soft
supersymmetry-breaking scalar contributions to the Higgs masses are allowed
to differ by the same amount from those of the squarks and sleptons at
the GUT scale, the so-called 
non-universal Higgs model 1 (NUHM1)~\cite{baer3,baer4,Ellis:2008eu}.

\begin{figure*}[htb!]
\begin{center}
\begin{picture}(300,400)
  \put(  0,   220){ \resizebox{10.5cm}{!}{\includegraphics{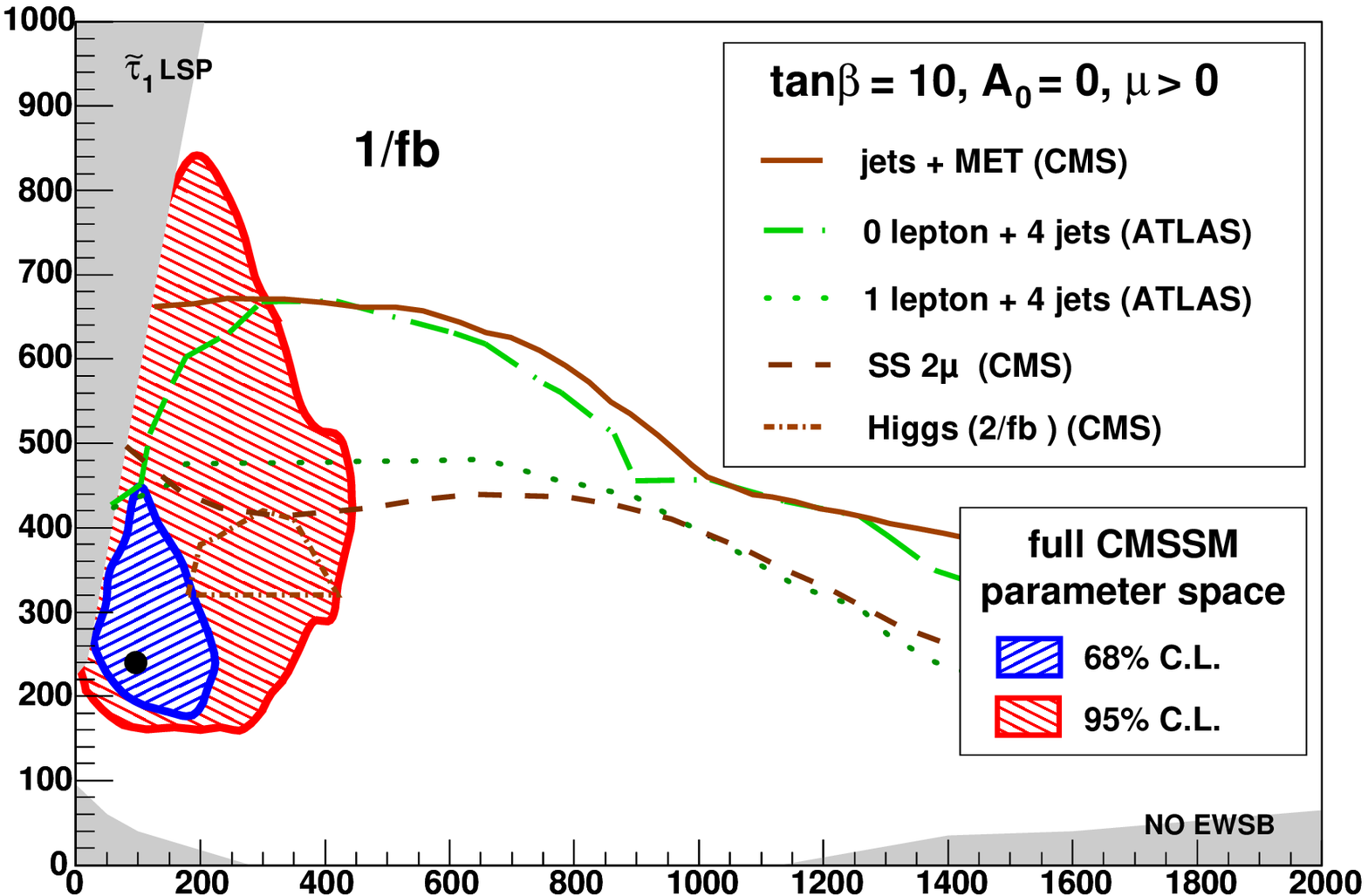}}}
  \put(250,   205){$m_0$ [GeV]}
  \put( -5,   360){\begin{rotate}{90}$m_{1/2}$ [GeV]\end{rotate}}
  \put(  0,     0){ \resizebox{10.5cm}{!}{\includegraphics{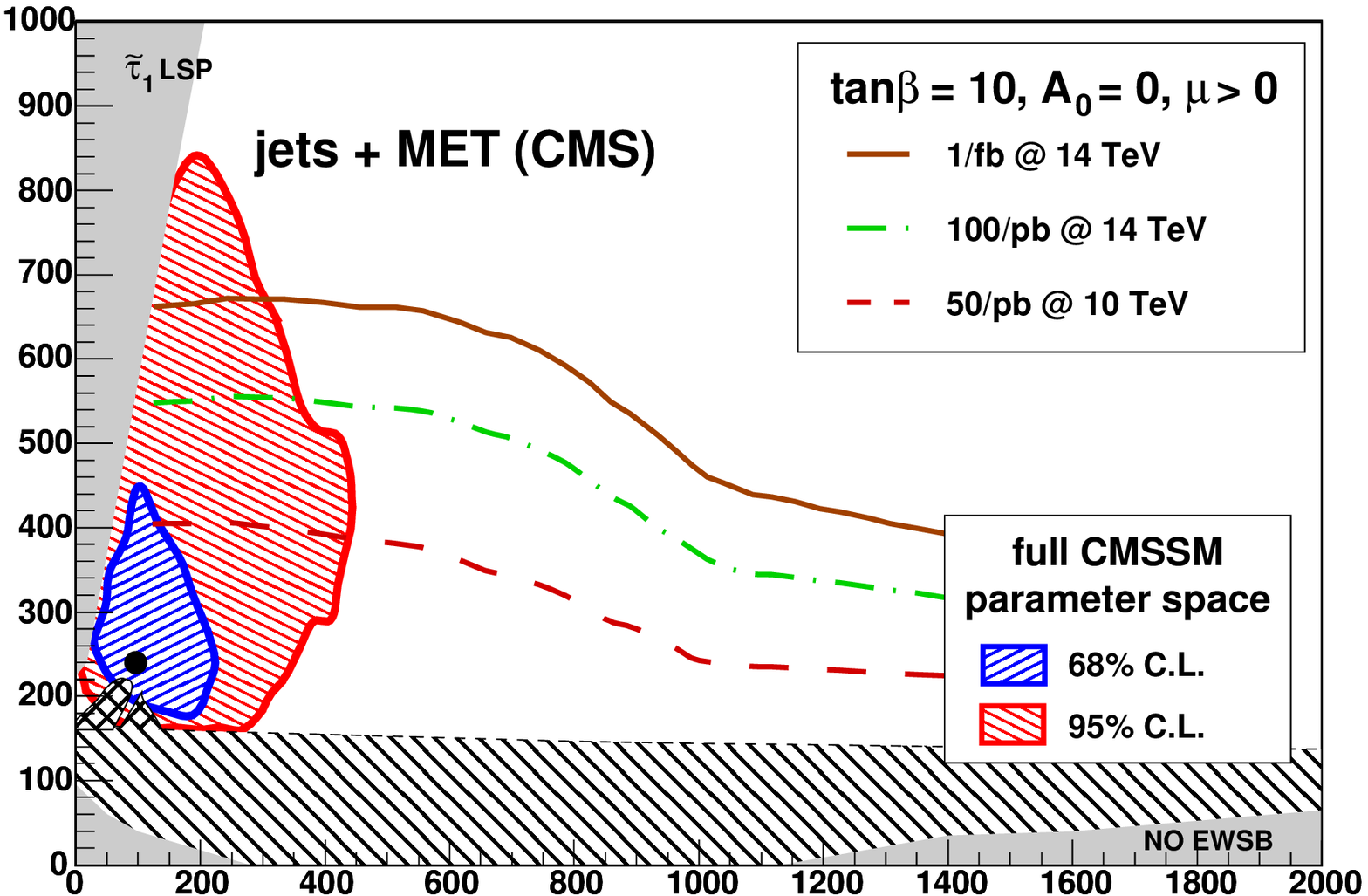}}}
  \put(250,   -15){$m_0$ [GeV]}
  \put( -5,   140){\begin{rotate}{90}$m_{1/2}$ [GeV]\end{rotate}}
\end{picture}
\end{center}
\vspace{-0.5em}
\caption {The $(m_0, m_{1/2})$ plane in the CMSSM for $\tan\beta = 10$
  and $A_0 = 0$, as in Fig.~\protect\ref{fig:contours}, 
  overlaid with  the 68\% and 95\% probability contours for the NUHM1.
  Upper plot: Some $5\,\sigma$ discovery contours at ATLAS and CMS with
  1~fb$^{-1}$ at 14~TeV, and the contour for the $5\,\sigma$ discovery
  of the Higgs boson in sparticle decays with 2~fb$^{-1}$ at 14~TeV in CMS.
  Lower plot: The $5\,\sigma$ discovery contours for jet + missing $E_T$ events 
  at CMS with 1~fb$^{-1}$ at 14~TeV, 100~pb$^{-1}$ at 14~TeV and
  50~pb$^{-1}$ at 10~TeV centre-of-mass energy.}
\label{fig:NUHM1}
\end{figure*}

Overall, it is encouraging that the general sizes of the 68\% and 95\%
C.L.\ regions are similar to those in the CMSSM, as shown in
Fig.~\ref{fig:NUHM1}, though the 68\% C.L.\ region together with the
best-fit point are shifted to lower $m_{1/2}$,
and the 95\% C.L.\ region is more elongated in $m_{1/2}$. 
As in the case of the CMSSM, SUSY could be
discovered over all of the 68\% C.L.\ region with 100~pb$^{-1}$ of
integrated luminosity at 14~TeV in a single
experiment, and even 50~pb$^{-1}$ of  integrated
luminosity at 10~TeV would cover most of it. As in the CMSSM,
not all of the NUHM1 95\% C.L.\ region would be covered by the LHC
with 1~fb$^{-1}$ of integrated luminosity at 14~TeV, whereas
the same-sign dilepton search would cover all the 68\% C.L.\ region
in the NUHM1. There are
differences between the shapes of the preferred regions in the CMSSM and
the NUHM1, particularly at low $m_{1/2}$.  This reflects the fact that
the $\Omega_{\rm CDM}$ constraint 
can be obeyed away from the coannihilation strip at larger values of $m_0$, if
$m_\chi \sim m_{H/A}/2$. This freedom can then be exploited to relax the
slight  tension induced by $\mh$ which arises in the CMSSM.

The spectrum at the best-fit NUHM1 point is shown in the right panel of
Fig.~\ref{fig:spectra}. This point
has $m_0 = 100$~GeV, $m_{1/2} = 240$~GeV, 
$A_0 = -930$~GeV, $\tan \beta = 7$, $m_H^2 = -6.9 \times 10^5$~GeV$^2$ 
and $\mu = 870$~GeV, yielding $\chi^2 = 18.0$ (39\% 
probability) 
and $\mh = 118$~GeV. The best-fit values of
$m_{1/2}$ and $\tan \beta$ are somewhat lower than those in the CMSSM,
whereas the value of $m_0$ is somewhat higher.
The overall value of $\chi^2$ is also somewhat
lower than in the CMSSM, reflecting the relaxation of the slight 
tension in the value of $\mh$
that is possible when the Higgs masses are allowed to become non-universal.
Comparing with the 
best-fit CMSSM spectrum, we see that the masses of the sleptons and
squarks are quite similar, as are the masses of the lighter neutralinos
and chargino. However, the splitting between the stop mass eigenstates
is larger --- reflecting the larger value of $|A_0|$,  the heavier
neutralinos and chargino are much heavier --- reflecting the larger
value of $\mu$, the best-fit value of $\mh$ lies comfortably above the
LEP lower limit, and the heavier Higgs bosons are lighter than in the
CMSSM --- reflecting the extra freedom conferred by the non-universality
in the NUHM1. The lower values of the heavier Higgs masses compensate
other SUSY contributions to \bsg, and offer better prospects for
detection at the LHC than those offered by the CMSSM.

\section{Conclusion and Outlook}

Making a probabilistic analysis using a MCMC technique, 
we have presented in this paper the regions preferred in the CMSSM and
the NUHM1 parameter spaces at the 68\% and 95\% C.L., as well as the
spectra at the best-fit points, in the light of the present direct and
indirect constraints on the models' parameters. Particularly important
roles are played by \gmt\ and \bsg, and we have analyzed the ways in
which effects of these constraints vary with the sizes of their 
theoretical uncertainties and experimental errors. 
We have quantified how strengthening (or relaxing) either
of these constraints would reduce (or expand) considerably the preferred
regions in the CMSSM $(m_0, m_{1/2})$ plane. We have also studied the
impact of the constraint on the cold dark matter density imposed by WMAP.
We find that the results for the best-fit points are remarkably robust
with respect to imposing or dropping the constraint on the cold dark
matter density.
Perhaps surprisingly, we find that this constraint 
does not restrict significantly most
two-dimensional projections of the preferred region in the CMSSM
parameter space. Encouragingly, we find that the preferred regions in the
NUHM1 are quite similar to those in the CMSSM.

The 95\% exclusion regions in the $(m_0, m_{1/2})$ plane extend
significantly further than the discovery regions shown above.
Therefore, if SUSY were to be excluded at the LHC
with 1~fb$^{-1}$ (100~pb$^{-1}$) of integrated luminosity at 14~TeV, the
95\% (68\%) C.L.\ regions in both the CMSSM and the
NUHM1 would be ruled out. On the other hand,
SUSY could be discovered at the 5~$\sigma$ level at the LHC with 1~fb$^{-1}$ 
of integrated luminosity at 14~TeV in a single experiment over most of
the 95\% C.L.\ regions in the $(m_0, m_{1/2})$ planes of the CMSSM
and the NUHM1. Only the
highest $m_{1/2}$ values would require a larger integrated luminosity,
or the combination of data from both ATLAS and CMS.
Indeed, SUSY could be discovered over all of the 68\% C.L.\ regions 
in both the CMSSM and the NUHM1 with just
100~pb$^{-1}$ of integrated luminosity at 14~TeV, and even 50~pb$^{-1}$ of 
(good-quality) data at 10~TeV would offer significant prospects for SUSY
detection. The same-sign dilepton search would cover most (all) of the
68\% C.L.\ region in the CMSSM (NUHM1). 
If Nature were to choose the best-fit CMSSM point, a
measurement of the same-sign dilepton endpoint would impose a strong
constraint on the SUSY spectrum.

One way or the other, there are good prospects that the initial runs of the
LHC will determine the fate of many speculations about the relevance of
low-energy SUSY to particle physics.

\section*{Acknowledgements}

We thank A.M.~Weber for collaboration in the early stages of this work.
This work was supported in part by the European Community's Marie-Curie
Research Training Network under contracts MRTN-CT-2006-035505
`Tools and Precision Calculations for Physics Discoveries at Colliders'
and MRTN-CT-2006-035482 `FLAVIAnet', and by the Spanish MEC and FEDER under 
grant FPA2005-01678. The work of S.H. was supported 
in part by CICYT (grant FPA~2007--66387), and
the work of K.A.O. was supported in part
by DOE grant DE--FG02--94ER--40823 at the University of Minnesota.

\newpage
\pagebreak



\end{document}